\documentclass{ametsocV6.1}

\usepackage[number-unit-product=~, inter-unit-product=~, list-final-separator=\text{, and }, range-phrase=\text{ to }, range-units=single, list-units=single]{siunitx}

\def\be{\begin{equation}}
\def\ee{\end{equation}}
\def\ba{\begin{eqnarray}}
\def\ea{\end{eqnarray}}

\renewcommand{\vec}[1]{\boldsymbol{#1}}

\title{Seasonality and spatial dependence of meso- and submesoscale\\
ocean currents from along-track satellite altimetry}

\authors{Albion Lawrence\aff{a}\correspondingauthor{Albion Lawrence, albion@brandeis.edu} and J\"orn Callies,\aff{b}}

\affiliation{\aff{a}{Brandeis University, Waltham, MA, USA}\\ \aff{b}{California Institute of Technology, Pasadena, CA, USA}}

\abstract{Along-track wavenumber spectral densities of sea surface height (SSH) are estimated from Jason-2 altimetry data as a function of spatial location and calendar month, to understand the seasonality of meso- and submesoscale balanced dynamics across the global ocean.
Regions with significant mode-1 and mode-2 baroclinic tides are rejected, restricting the analysis to the extratropics. 
Where balanced motion dominates, the SSH 
spectral density is averaged over all pass segments in a region 
for each calendar month, and is fit to a 4-parameter model consisting of a flat plateau at low wavenumbers, a transition at wavenumber $k_0$ to a red power law spectrum $k^{-s}$, and a white spectrum at high wavenumbers that models the altimeter noise. The monthly time series of the model parameters are compared to the evolution of the mixed layer. The annual mode of the spectral slope $s$ reaches a minimum after the mixed layer deepens, and the annual mode of the bandpassed kinetic energy in the ranges $[2k_0,4 k_0]$ and $[k_0,2 k_0]$ peak $\sim$2 and $\sim$4 months, respectively, after the maximum of the annual mode of the mixed layer depth. This analysis 
is consistent with an energization of the submesoscale by a winter mixed layer instability followed by an inverse cascade to the mesoscale, in agreement with prior modeling studies and {\it in situ} measurements. These results are compared to prior modeling, {\it in situ}, and satellite investigations of specific regions, and are broadly consistent with them within measurement uncertainties.}

\begin{document}

\maketitle

\statement
This paper uses satellite observations to understand the source of ocean dynamics at the \SIrange{1}{100}{km} scales at which vertical motion becomes important and which are thus relevant for biology and for the exchange of heat and carbon with the atmosphere. The observations are consistent with a seasonal variation of dynamics at these scales, predicted by a specific theory of upper-ocean turbulence and confirmed by modeling studies and regional observations. We update prior satellite-based studies by excluding regions with competing effects, by our treatment of the noise, and by our characterization of the seasonality. This work provides a template for analyzing data from the upcoming Surface Water and Ocean Topography (SWOT) satellite. 

\section{Introduction}

Over the past decade, {\it in situ}\ observations \citep{Callies2015-bu,Thompson2016-vq,Damerell2016-hc,Erickson2018-xk,Erickson2020-ij} and modeling \citep{Sasaki2014-cq,Qiu2014-hn,Rocha2016-kd,Chereskin2019-dd} have combined to provide a picture of the energization of submesoscale (\SIrange{1}{100}{km}) ocean currents via mixed instabilities (MLIs), triggered by wintertime deepening of mixed layers combined with lateral buoyancy gradients that provide available potential energy \citep{Boccaletti2007-oz,Fox-Kemper2008-xy}. The observational signals include a wintertime shallowing of the spectrum of balanced kinetic energy (KE) in the \SIrange{10}{100}{km} range; modeling reveals peaks in bandpassed KE with timing relative to that of the mixed layer depth (MLD) that increases with spatial scale, indicating an upscale cascade in this range of energy injected by MLI.

In this work, we use satellite altimetry observations from the Jason-2 mission to assess how widespread this phenomenon is across the global extratropical ocean. While current altimetry, which resolves balanced motion at wavelengths greater than about \SIrange{50}{70}{km} cannot capture the scales at which MLI injects energy, there is a long enough record to detect the fingerprint of MLI and the subsequent inverse cascade in the seasonal characteristics of the observable balanced motion. While there have been several recent altimetry-based studies of balanced motion at these scales \citep[e.g.][]{Xu2011-ym,Xu2012-cm,Vergara2019-kz}, our analysis has a number of distinct features:
\begin{itemize}
    \item We reject regions with substantial baroclinic mode-1 and mode-2 diurnal tides. Theese tides can project onto the SSH signal seen by the altimeter and can be dominant precisely in the wavenumber range we are interested in. Furthermore, there is evidence that this signal has a seasonality that competes with the effects of interest \citep{Rocha2016-kd,Qiu2018-ra,Chereskin2019-dd}. This restricts our observations to the extratropics and regions where balanced flow is strong compared to the internal tides \citep[cf.][]{Callies2019-yf}.
    \item Satellite observations are limited by white noise that dominates at wavelengths lower than \SIrange{50}{70}{km}. Rather than simply measuring the amplitude and subtracting it out, we include the noise in our model of the altimeter spectrum. We are careful to check that the seasonality of the noise does not contaminate the seasonality of the estimated balanced motion. One result of this approach is that spectral slopes diagnosed here are closer to those found from modeling and from {\it in situ} measurements than in previous altimetry-based studies.
    \item To isolate the annual component of the seasonal signal, we extract the annual Fourier component of a monthly time series of the balanced motion relative to that of the MLD, the latter computed from the ECCO-4 climatology. This is distinct from the standard strategy of binning spectra over fixed 3-month periods. We will find temporal structures on the order of 2~months, which this latter approach can miss.
\end{itemize}
The upshot is that the surmised energization of submesoscales by a wintertime MLI is broadly consistent with data across the global ocean, with some notable exceptions -- particularly the Southern Ocean south and west of the African continent -- worthy of further investigation. 

The remainder of our paper is organized as follows. In Section~\ref{S:2}, we describe the data set and methods used to analyze the seasonality of balanced motion, motivated by the spectra from a well-studied region in the Northwest Atlantic near the Gulf Stream. In Section~\ref{S:3}, we display the results of our analysis for the global ocean. In Section~\ref{S:4}, we revisit a number of specific regions that have been previously examined via modeling studies, {\it in situ} measurements, and satellite altimetry. In Section~\ref{S:5}, we present our conclusions and suggest directions for future work. The Appendices provide additional technical details of our analysis.

\section{Data and Methods}\label{S:2}

We use Jason-2 along-track data from the years \SIrange{2008}{2016}. Following the procedure described in Appendix~A, we estimate one-dimensional (along-track) power spectra for the SSH signal between approximately \SIlist{12;900}{km} wavelengths, the lower limit being the Nyquist wavelength and the upper limit corresponding to the length of the segments we allow. The spectra are averaged over overlapping $\ang{8}\times \ang{8}$ boxes, whose centers are spaced \ang{4} apart in latitude and longitude, between \ang{58}S and \ang{58}N. As the spectra are calculated after multiplying the SSH along each pass segment by a window function, these partially overlapping squares should provide approximately independent data sets.

\begin{figure}[t]
  \noindent\includegraphics[scale=0.6]{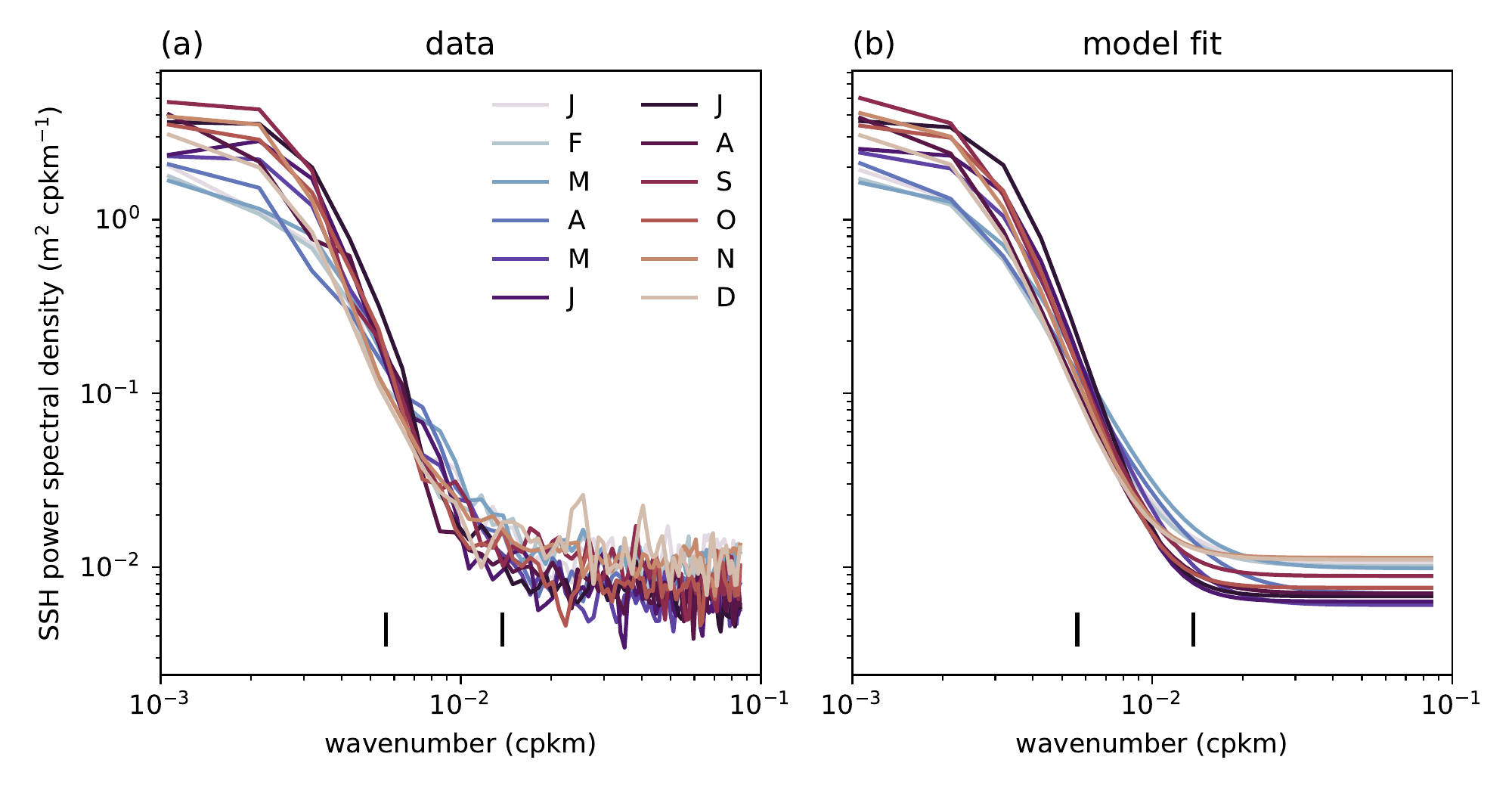}
  \caption{(a)~Along-track SSH power spectra in an $\ang{8} \times \ang{8}$ box centered at \ang{32}N \ang{68}W, averaged over all passes through the box in each calendar month. (b)~Model spectra from fitting \eqref{eq:tidefreemodel} to the measured power spectrum in each month. The vertical lines are the wavenumbers of the semidiurnal baroclinic mode-1 and mode-2 tides. In this region, the tidal signal is negligible compared to that of the balanced flow.}
  \label{f:Gulf_monthly_spectra}
\end{figure}

An example of the spectra binned by month can be seen in Fig.~\ref{f:Gulf_monthly_spectra}a. This is taken from an $\ang{8}\times\ang{8}$ box centered at \ang{32}N \ang{68}W, a region of the western Atlantic just south of the Gulf Stream.\footnote{In fact, this patch is centered on the corners of four adjacent patches in our global map; however, it is a useful patch for benchmarking our analysis against the {\it in situ} data we discuss in Section~\ref{S:4}.\ref{ss:regional}.\ref{sss:gs}} This region has energetic mesoscale turbulence and a strong seasonal cycle in the MLD, and it includes the location of the Oleander data discussed in \citet{Callies2015-bu}. The spectra in Fig.~\ref{f:Gulf_monthly_spectra}a have a simple qualitative form: they roll off from a low-wavenumber plateau to a red spectrum, approximately following a power law, before transitioning to a white spectrum at \SIrange{50}{100}{km} that reflects the altimeter noise \citep{Xu2011-ym,Xu2012-cm,Dibarboure2014-jd}. The scale of this latter transition, which is determined both by the strength of the noise and the strength of the balanced signal, sets the effective resolution of the altimeter.

Fig.~\ref{f:Gulf_monthly_spectra}a already illustrates some basic features of the time dependence in these spectra. As noted in {\it in situ}\ measurements within this region \citep{Callies2015-bu}, the spectral roll-off between the low-wavenumber plateau and the high-wavenumber noise floor is notably gentler in the winter months. This is consistent with the injection of energy at short distances (the \SIrange{1}{10}{km} range) in the winter months, followed by an inverse cascade into the range observable by the altimeter \citep[cf.][]{Sasaki2014-cq}.

Our goal is to quantitatively characterize the seasonality and its statistical significance across the global ocean and to disentangle it from the seasonality in the altimeter noise, which is also visible in Fig.~\ref{f:Gulf_monthly_spectra}a.

Our first step is to carefully delineate the regions in which we can reliably extract the sought-after information from the altimeter. In addition to rejecting regions with low data quality (see Appendix A), we must be careful of signals that compete with the balanced dynamics, specifically the semi-diurnal mode-1 and mode-2 baroclinic tides \citep[e.g.][]{Ray2016-pi,Callies2019-yf}. The tides project onto the SSH precisely in the wavenumbers of interest, and can be a significant component of the spectrum. Furthermore, the tidal signal has been seen in modeling studies to have a seasonality of its own, with phase opposite to that of the balanced flows \citep{Rocha2016-kd,Qiu2018-ra,Lahaye2019-xm,Chereskin2019-dd}. Thus, we reject regions with a significant tidal signal (more precisely, we reject regions with a significant mode-1 signal), as described in Appendix C.  A consequence of this criterion is that our analysis is restricted to the extratropics. This rejection of regions with significant tidal signal in the wavenumbers of interest is one significant difference between our work and prior global studies of SSH seasonality at these scales.

\begin{figure}[t]
  \noindent\includegraphics[scale=0.55]{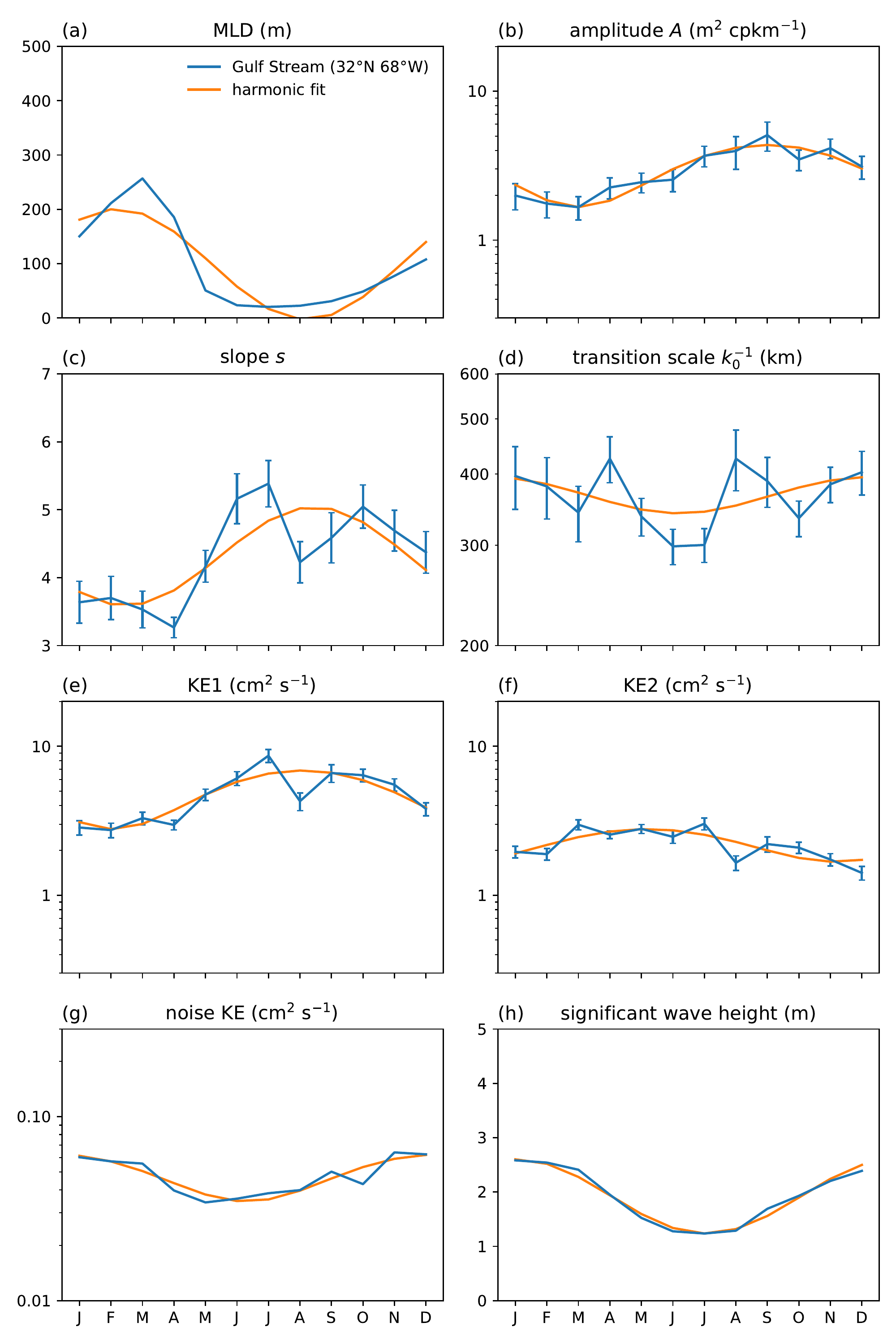}
  \caption{Monthly time series of model parameters from the fit of \eqref{eq:tidefreemodel} to the SSH power spectra from an $\ang{8} \times \ang{8}$ region centered at \ang{32}N \ang{68}W. (a)~The mixed layer depth from ECCO-4 climatology. (b)~The plateau amplitude~$A$. (c)~The spectral slope~$s$. (d)~The transition scale~$k_0^{-1}$. (e)~The bandpassed kinetic energy KE1 (\SIrange{176}{351}{km}). (f) The bandpassed kinetic energy KE2 (\SIrange{88}{176}{km}). (g)~The altimeter noise interpreted as balanced KE. (h)~The significant wave height from ERA5. The computation of $A$,$s$,$k_0$, the noise, and the displayed uncertainies, are described in Appendix~B.}
  \label{f:gulf_parameters}
\end{figure}
\clearpage

For those grid boxes that survive tidal rejection, we then compute the power spectrum averaged over all pass segments and cycles which pass our data quality filter for a given calendar month and fit them to a model of the balanced flows and altimeter noise without tides \citep[cf.][]{Callies2019-yf}:
\be
	P(k; \phi, \lambda, t) = \frac{A(\phi, \lambda, t)}{1 + [k/k_0(\phi, \lambda, t)]^{s(\phi, \lambda, t)}} + N(\phi, \lambda, t). \label{eq:tidefreemodel}
\ee
Here, $t$ refers to the calendar month 
and $\phi$ and $\lambda$ to the latitude and longitude of the center of the grid box. The details of our fitting procedure are given in Appendix B, which also describes the computation of error bars $\pm \sigma$ for our estimate of each of the parameters shown; the uncertainty in the phase and amplitude of the annual mode are computed via the standard formula for propagation of errors.  We also extract an MLD $h(\phi, \lambda, t)$ in each grid box by averaging the ECCO-4 climatology over said grid box. In Fig.~\ref{f:Gulf_monthly_spectra}b, we show the model SSH power spectrum $P(k, \phi, \lambda, t)$ fit to the data shown in Fig.~\ref{f:Gulf_monthly_spectra}a, for which $\phi = \text{\ang{32}N}$ and $\lambda = \text{\ang{68}W}$. The qualitative seasonality matches that seen in the raw data. We can see this even more precisely in Fig.~\ref{f:gulf_parameters}a--d, displaying the parameters $A$, $k_0$, $s$, $N$ as well as the MLD by month. Note that, consistent with \citet{Callies2015-bu}, the slope~$s$ is lowest (and the roll-off is gentlest) when the mixed layer is deepest. Note also that the values of $s$ in the summer and winter months are broadly consistent with that work; they are between \numlist{3.5;4.0} in the winter months (corresponding to \numlist{1.5;2.0} for KE spectra), and between \numlist{5.0;5.5} in the summer months (corresponding to \numlist{3.0;3.5} for KE spectra).  We should be careful, however, in interpreting the value of~$s$. The apparent power law behavior in our spectra lasts for less than a decade in wavenumber between $k_0$ and the noise floor; the error bars in $s$ are $\sim$0.5; $s$~has considerable variations month by month; and, as we will see, the data are consistent with periodic rather than constant injection of energy via MLI, complicating standard inertial-range explanations. That being said, $s$ can still be taken as a measure of the steepness of the spectrum between $k_0$ and the noise floor.

At the submesoscale distances we are probing here, the unbalanced internal gravity wave continuum \citep{Garrett1972-va,Garrett1975-md,Garrett1979-li} is also active and projects onto SSH, with a smaller spectral slope than the balanced motion. We will ignore these motions as we believe the white altimeter noise dominates at scales where this would be an issue \citep{Callies2019-yf}.

The seasonality of the slope~$s$ is a coarse measure of the inverse cascade that is triggered by MLI. To get a more refined picture, we use the model \eqref{eq:tidefreemodel} to compute the bandpassed kinetic energy of the balanced motion in two adjacent wavenumber bands, following the treatment in \citet{Wortham2014-ze,Wortham2014-ob}. We define the quantity ``KE1'' as the kinetic energy integrated in the band $[k^{ann}_0:2k^{ann}_0]$,  and ``KE2'' as the kinetic energy integrated over the band $[2k^{ann}_0:4k^{ann}_0]$, where $k^{ann}_0$ is computed by fitting \eqref{eq:tidefreemodel} to the annually averaged spectrum.\footnote{This is not an ideal definition of wavenumber bins for studying the effects of MLI, which takes place at much shorter scales. However, these bins do well in capturing the range of balanced motion, below the mesoscale eddy scale, which is available to the altimeter.} We expect that an inverse cascade triggered by MLI at $\lesssim$\SI{10}{km} scales as the mixed layer deepens would lead to KE2 peaking somewhat after the month of maximum mixed layer depth, and KE1 peaking somewhat after that, similar to the model results described by \cite{Sasaki2014-cq}. We can see this behavior explicitly in Fig.~\ref{f:gulf_parameters}e--f for the Gulf Stream region we have used as an example in this section.

There is also seasonality in $A$ and $k_0$, as seen in Fig.~\ref{f:gulf_parameters}e--f. These parameters, together with the slope, determine KE1 and KE2. In different combinations $A$, $k_0$, and $s$ also characterize mesoscale phenomena such as the location and height of the peak in the kinetic energy \citep[e.g.][]{Stammer1997-fp,Tulloch2011-dc}. 

As we can see from Fig.~\ref{f:gulf_parameters}g, the altimeter noise has its own seasonality, tracking that of the significant wave height (SWH) (Fig.~\ref{f:gulf_parameters}h) computed from ERA5 reanalysis \citep{Hersbach:2020}. However, in this patch, the bandpassed kinetic energies have a distinct seasonality from the observed noise. We will find that this is generally true, giving us some confidence that the seasonality we are observing is a genuine feature of the balanced dynamics, rather than being an imprint of the altimeter noise.

The time series we have examined show definite seasonality correlated with the seasonality in the MLD. To compare these time series across the global ocean, we would like to use more precise quantities. One standard technique is to further bin the spectra into three-month winter and summer ``seasons,'' defined across the Northern or Southern Hemisphere. We adopt a different tack for two reasons. First, there is important temporal structure in the timing difference between the annual mode of various observables, at the level of one to two months. Second, there are temporal variations on many time scales; since we expect seasonality to be driven by a strong annual cycle, we are most interested in the responses of the annual mode of the balanced motion.  

\begin{figure}[t]
  \noindent\includegraphics[scale=0.6]{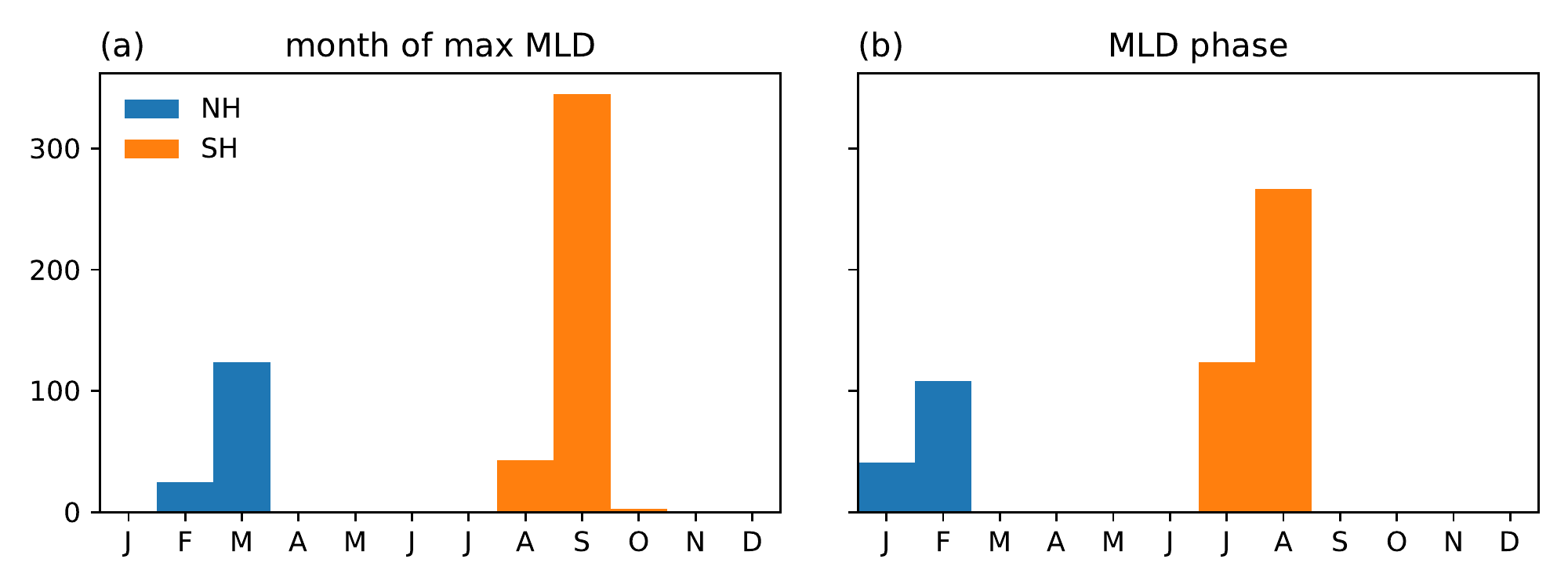}
  \caption{(a)~Histogram of the calendar months of the maximum of mixed layer depth (MLD) in each grid box in which SSH power spectra are produced, computed from ECCO-4 climatology. Blue bars are for points in the Northern Hemisphere (NH), and orange for points in the Southern Hemisphere (SH). (b)~Histogram of the calendar month of the maximum of the ``annual mode'', the Fourier component of the monthly time series of the MLD with an annual period.}
  \label{f:mld_phase}
\end{figure}

To get more refined temporal information, then, for each parameter $q$ running over the parameters in \eqref{eq:tidefreemodel}, the MLD, and the bandpassed energies KE1 and KE2, we compute the amplitude $|\hat q|$ and the phase $\theta_q$ (in months) of the annual mode:
\be
	\sum_{t = 1}^{12} e^{2\pi i (t - 1)/12} q(t) = |\hat q| e^{2\pi i\theta_q/12}
\ee
(where $t = 1$ corresponds to January). We determine the confidence limits of these signals by computing the log likelihood of the annual mode being zero.
In the region shown in Fig.~\ref{f:Gulf_monthly_spectra} and~\ref{f:gulf_parameters}, $|\hat q|$ is nonzero for all of the model parameters at greater than $99\%$ confidence. Under the assumption that the annual modes of the parameters in \eqref{eq:tidefreemodel} respond directly to the annual mode of the MLD, we will focus in particular on the relation of the phases of the annual modes of $s$, KE1, and KE2 to the phase of the annual mode of the MLD, as a measure of the time delay between the deepening of the mixed layer and the response at scales larger than the injection scale.  In the Gulf Stream region we are using as an example, $\theta_{-s} - \theta_h = 0.2 \pm 0.5$~months, $\theta_\mathrm{KE2} - \theta_h = 3.0 \pm 0.4$~months, and $\theta_\mathrm{KE1} - \theta_h = 5.9 \pm 0.3$~months. Note, as we can see from Fig.~\ref{f:mld_phase}\ that the maximum of the annual mode generally peaks a month earlier than the maximum of the MLD itself and is slightly less peaked on the modal value than the maximum of the MLD. 

We close this discussion of the data and our methods with two comments on the altimeter noise. First, a significant difference of our analysis from prior published work is that we explicitly include the amplitude of the noise in our model of the SSH wavenumber spectral density, rather than simply estimating the noise from the high-wavenumber signal and subtracting it. The uncertainties in our estimation of the noise are thus naturally taken into account in our uncertainties in estimating the parameters of our model for the balanced spectrum. Second, an open possibility is that spatial correlations in the noise could redden the noise in the region we attribute to the balanced flows \citep{Xu2012-cm,Dibarboure2014-jd}, so that we overestimate the latter. If the noise rises sufficiently slowly compared to the apparent power law of the balanced motion, this will not have too large of an effect. To estimate this, we followed the analysis of \citet{Xu2012-cm} and compute the spectrum of the difference between Jason-1 and Jason-2 data during the cross-calibration period (July 4, 2008 to Jan 26, 2009), as a measure of the altimeter noise. At the largest length scales captured in the spectra, the signal-to-noise ratio ranges from \numrange{e2}{e6}. 

\section{Results}\label{S:3}

\subsection{SSH spectral slopes}

The SSH spectral slope for wavenumbers above $k_0$ is a subject of intense investigation, as this number is the output of several competing theories of upper-ocean turbulence and its activation. We thus open with a global analysis of the parameter $s$ analyzed with our own protocols, keeping in mind our caveats regarding the short wavenumber range over which this slope is measured. In Section~\ref{S:4}, we will compare these results to prior work.

\begin{figure}[t]
  \noindent\includegraphics[scale=0.6]{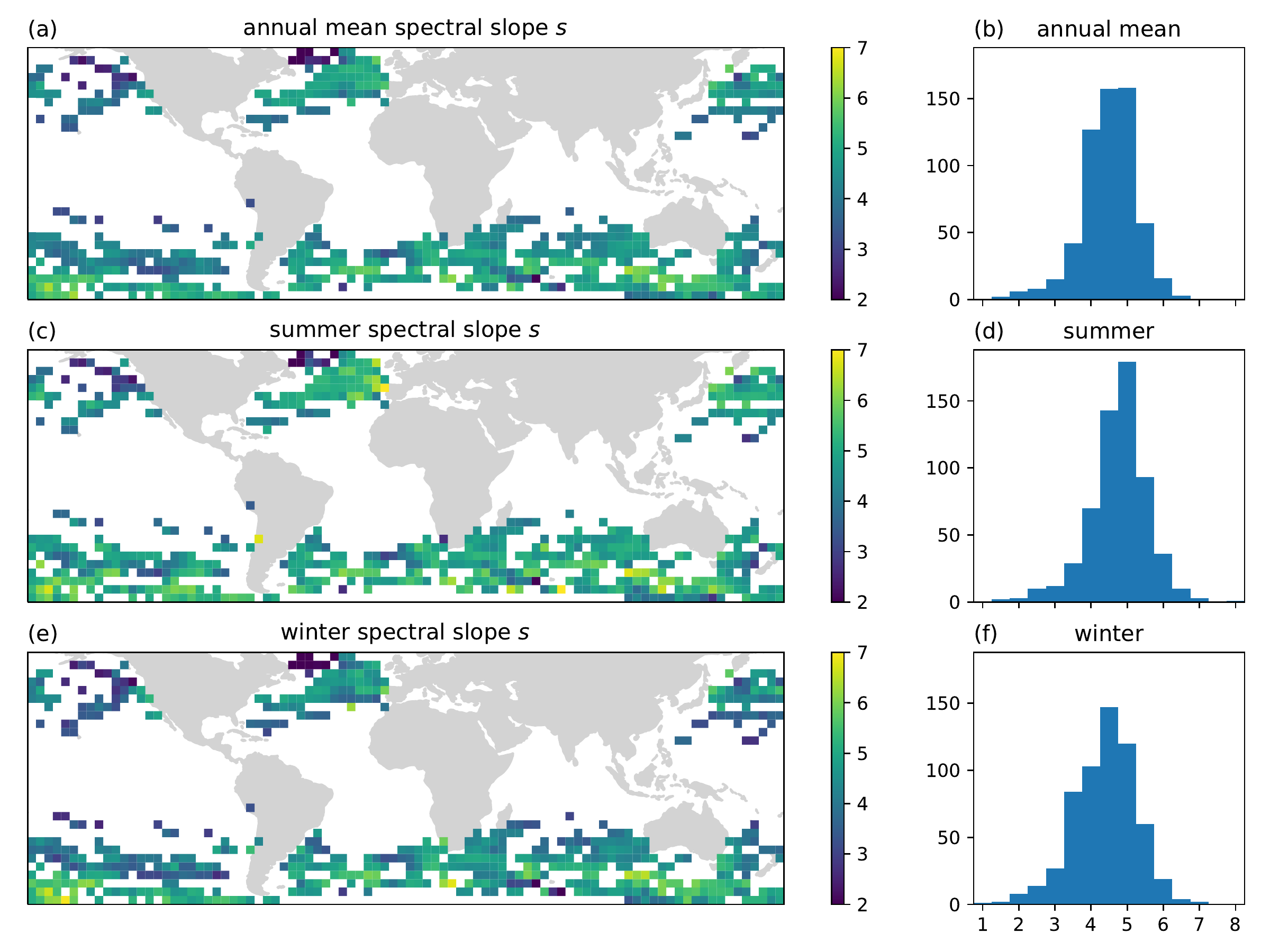}
  \caption{(a)~Global map of the SSH spectral slope~$s$ estimated by fitting the SSH power spectrum, averaged over the entire data set, to the model~\eqref{eq:tidefreemodel}. The standard deviation of~$s$ is typically on the order of~\num{.1}. (c)~Spectral slopes for SSH power spectra averaged over summer months (JAS in the Northern Hemisphere and JFM in the Southern Hemisphere). (e)~Spectral slopes for SSH power spectra averaged over winter months (JFM in the Northern Hemisphere and JAS in the Southern Hemisphere). (b,d,f)~Histograms of slopes mapped in (a,c,e), respectively.}\label{f:slope_annual_map}
\end{figure}

Figure~\ref{f:slope_annual_map}a maps $s$ for the spectra obtained by averaging over all calendar months. The errors in these values are of the order of $\sigma_s \sim 0.1$. Note that the slopes obtained here are considerably larger than those computed in \citet{Xu2012-cm} via a distinct algorithm for subtracting the noise and fitting the remaining spectrum (an issue we will discuss further in Section~\ref{S:4}). In our analysis, the resulting slopes cluster around $s = \numrange{4}{5}$, corresponding to KE spectra behaving as $k^{-2}$ to $k^{-3}$, and can be higher. 
We note that there are some regions, particularly the high-latitude Northwest Atlantic and Northeast Pacific, with low slopes. Closer inspection of the power spectra in these regions shows that there is no clear plateau at low wavenumbers, whereas we would expect $k_0$ to be larger at high latitudes, following the decrease in the deformation radius. We do not have an explanation; for the North Atlantic, large-scale variability of the Labrador Current and the effects of sea ice on the altimetry signal could play a role. 

As we can see from the example in Fig.~\ref{f:gulf_parameters}d, the slope can vary by $\gtrsim$1 as a function of calendar month, potentially reflecting time-varying injection of kinetic energy at scales below the altimeter's resolution. To get a global sense of this variation, in Figs.~\ref{f:slope_annual_map}b--c we plot the slopes for spectra averaged over three winter months (JFM in the Northern Hemisphere and JAS in the Southern Hemisphere); and three summer months (JAS in the Northern Hemisphere and JFM in the Southern Hemisphere).\footnote{These bins were chosen to match the bins used in \citet{Callies2015-bu} and match the seasonality in the MLD fairly well.} One can see from the color maps as well as the histogram a shift to gentler slopes in the winter. The mean winter slope is $s = 4.4$, and the mean summer slope is $s = 4.8$; the mean error is $\sim$0.2, although errors range from \numrange{.08}{1.0} in the winter, and \numrange{.08}{1.6} in the summer; the skewness of the winter slopes is $-0.4$ and that of the summer slopes is $-0.6$.

We could do a more refined analysis of these slope differences: for example, examining variance of the slope as a function of spatial location or examining the effect of moving the three-month windows. As we have seen for a specific example in Section~\ref{S:2}, however, interesting dynamics in the range $k > k_0$, characteristic of the inverse cascade following MLI, happen on the time scale of this window \citep[cf.][]{Sasaki2014-cq,Dong2020-ea}. We will therefore focus on the phase of the annual mode, measured in months, 
as outlined in Section~\ref{S:2}.

\subsection{Timing of balanced motion vs. MLD}

In Section~\ref{S:2}, we described the application of our analysis techniques via the specific example of a region near the Gulf Stream, which displays the seasonal behavior expected from a submesoscale MLI triggered in the winter months by a deepening mixed layer:
\begin{itemize}
    \item The annual mode of the slope parameter has a minimum, corresponding to a gentler slope, near the maximum of the annual mode of the MLD, due to an enhancement of submesoscale energy.
    \item The bandpassed kinetic energy in two adjacent wavenumber bins covering the red region of the SSH spectrum have annual modes with maxima that peak months after the peak of the annual mode of the MLD; and the peak in the bandpassed KE for adjacent wavenumber bands is later for lower wavenumbers, consistent with an inverse cascade from the submesoscale to the mesoscales. 
\end{itemize}
In this section we show results for the global ocean and provide evidence that this story dominates in the extratropical regions for which we have reliable data.

\begin{figure}[t]
  \noindent\includegraphics[scale=0.6]{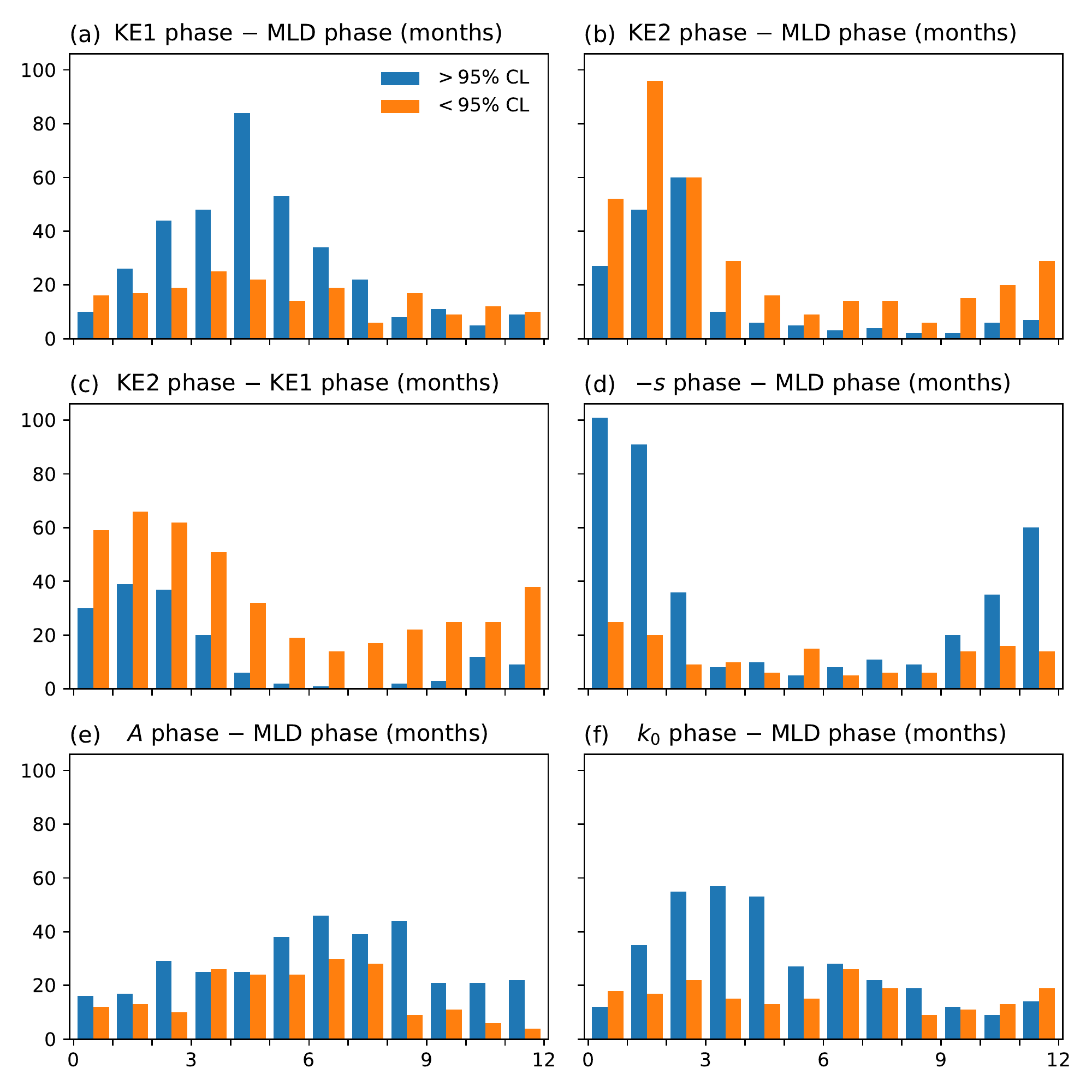}
  \caption{(a)~Histogram over $\ang{8}\times \ang{8}$ grid boxes of phase difference in months for the annual mode of the kinetic energy integrated over the range $[k_0,2k_0]$ (KE1), with respect to the phases of the annual modes of the mixed layer depth (MLD) computed from the ECCO-4 climatology. The KE is computed from the balanced component of the fit to (\ref{eq:tidefreemodel}). Blue corresponds to grid boxes for which the signal is nonzero at a $>$95\% confidence level; orange corresponds to the rest. (b)~The same for kinetic energy integrated between $[2k_0,4k_0]$ (KE2). (c)~Difference in phases for the annual modes of KE1 and KE2; blue corresponds to grid points for which both annual modes are nonzero at a $>$95\% confidence level. (d--f)~As in (a) and (b), but for (d)~the slope $-s$, (e)~the plateau amplitude~A, and (f)~the transition scale~$k_0$ derived from the model \eqref{eq:tidefreemodel}.}
  \label{f:seasonal_vs_mld}
\end{figure}

Figure~\ref{f:seasonal_vs_mld}a--d shows histograms of the phases (in months) of the annual components of the spectral slope $-s(\phi,\lambda,t)$ and the integrated kinetic energies KE1 and KE2 computed from the balanced component of the model~\eqref{eq:tidefreemodel}, relative to the phase of the climatological MLD. Each data point corresponds to a specific $\ang{8} \times \ang{8}$ box. We also plot histograms of the phase difference between KE1 and KE2. Note that we have separated out the data points with annual components that are nonzero at a $>$95\% confidence level; those which do not have statistically significant seasonality at this level show a comparatively flat distribution of phase differences with respect to the MLD, as may be expected. We checked the timing of the annual components of these observables for data points with nonzero annual modes in these observables between 95\% and 99\% confidence and found their measured behavior broadly similar to those with nonzero annual modes at $>$99\% confidence. 

For grid points in which the annual mode is statistically significant at $>$95\% confidence, we find that the slope becomes gentlest within one to two months of the maximum of the annual mode of the MLD, KE2 peaks within two to three months, while KE1 peaks within four to five months of the MLD, indicating that kinetic energy is moving upscale. The phase difference between KE1 and KE2 is generally on the order of one to three months. These results are consistent with an inverse cascade triggered by an instability that injects energy at scales below that at which the balanced motion is visible above the altimeter noise.

\begin{figure}[t]
  \noindent\includegraphics[scale=0.6]{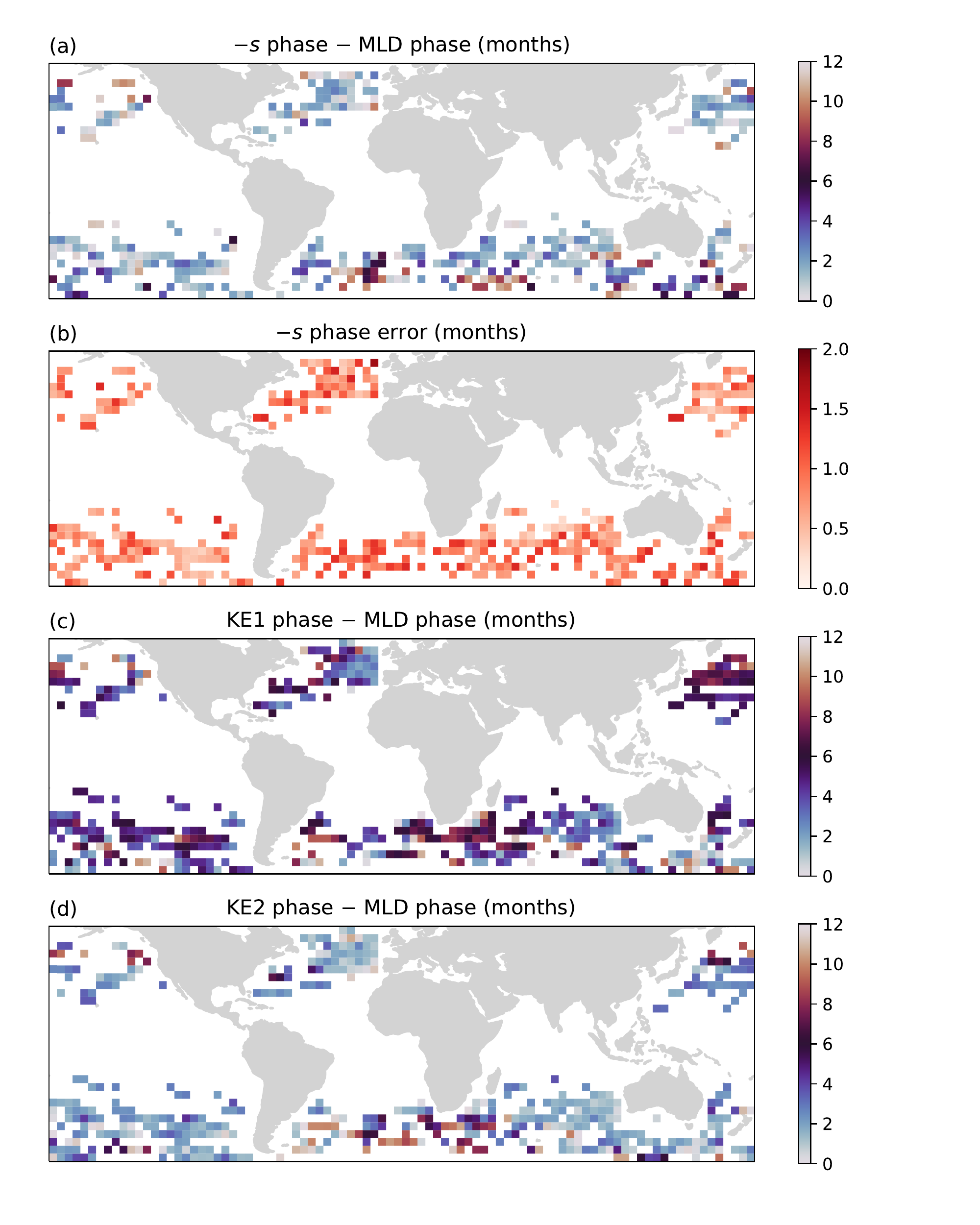}
  \caption{(a)~Global map of the phase in months of the annual mode of the SSH spectral slope $-s$ with respect to the phase of the annual mode of the mixed layer depth (MLD). The slope~$s$ is estimated by fitting \eqref{eq:tidefreemodel} to monthly spectra. Where the color is dark blue, the SSH power spectrum falls of the gentlest (is least red) just after the mixed layer is deepest. (b)~Map of errors in the phase of the annual mode of~$s$. (c)~Map of the phase of KE1, the kinetic energy integrated over $[k_0,2k_0]$. (d)~Map of the phase of KE2, the kinetic energy integrated over $[2k_0,4k_0]$.}
  \label{f:slopemap}
\end{figure}

One goal of this work is to understand the spatial dependence of mesoscale and submesoscale seasonality. In Fig.~\ref{f:slopemap}a we show a map of the phase of the annual mode of $-s(\phi,\lambda,t)$, measured with respect to the MLD phase as computed from ECCO-4, with the phase error in Fig.~\ref{f:slopemap}b. As we can see from the map, for most grid points passing our selection criteria, the dominant story is that the annual mode of the SSH power spectral slope is gentlest within about two months of when the annual mode of the MLD is maximal. There are regions where the phase difference between the minimum of the annual mode of the slope and the maximum of the annual mode of the MLD is large, particularly in the South Atlantic and South Indian; we believe these regions are worth more intensive study.

We can get more refined information about the inverse cascade by looking at the seasonality of dynamics in different wavenumber bins for $k > k_0$. In Fig.~\ref{f:slopemap}c--d, we map the phases relative to those of the climatological MLD of the bandpassed kinetic energies KE1 and KE2 computed from the balanced component of the model~\eqref{eq:tidefreemodel} for grid points in which the signals are nonzero at $>$95\% confidence. Again, the results are broadly in agreement with expectations from submesoscale energization via MLI, followed by an inverse cascade. Looking at KE2 in particular in Fig.~\ref{f:slopemap}d, there is a one- to two-month difference between the phases of the annual mode of the KE2 signal and the annual mode of the MLD. However, there is a noticeable difference in the South Atlantic and in the Southern Ocean south of the African continent. There is a hint of increased uncertainty in the phase error in these regions, but not enough to explain the substantial lag in the timing of KE2 relative to our baseline expectations from MLI-induced energization of the submesoscale.

\begin{figure}[t]
  \noindent\includegraphics[scale=0.6]{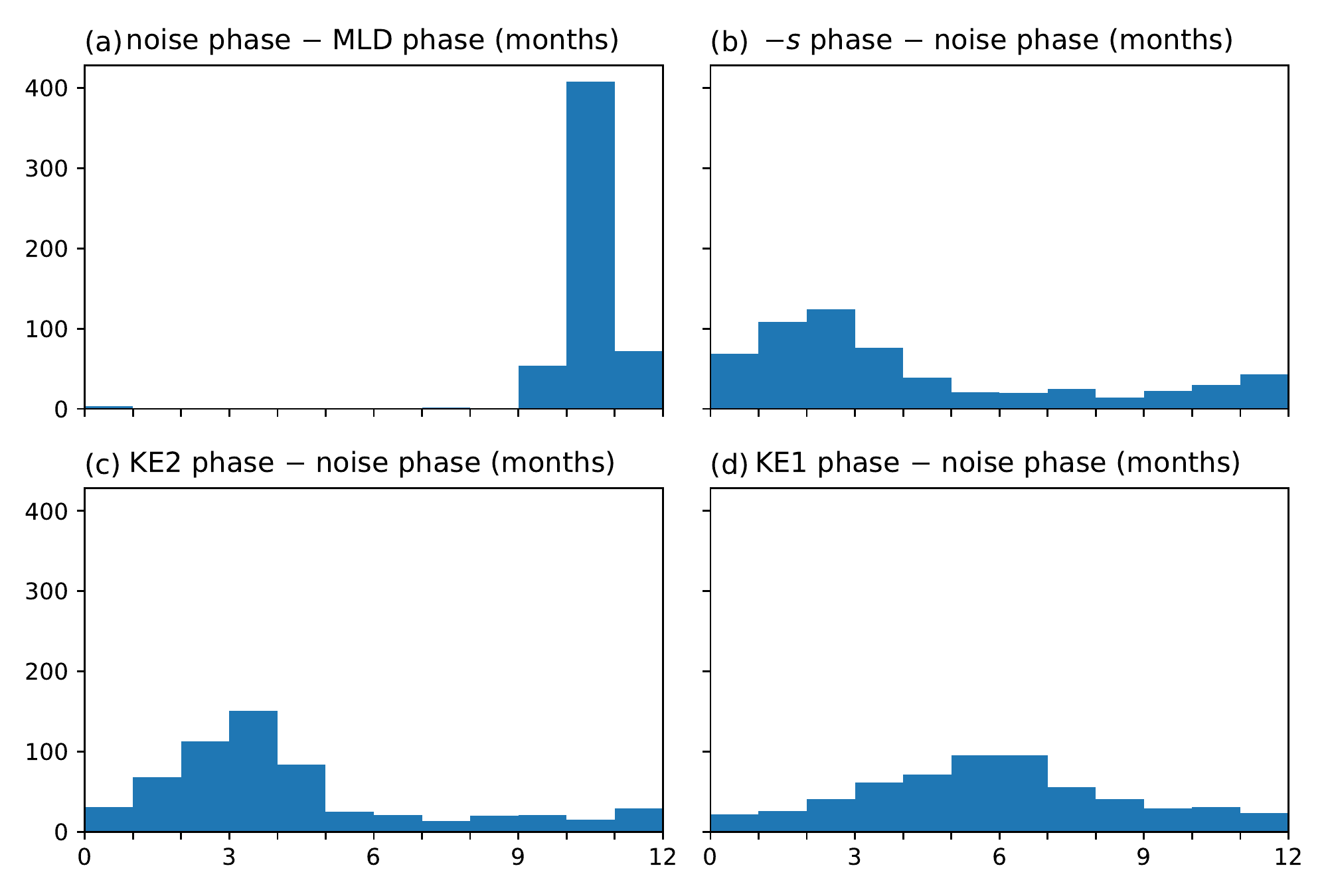}
  \caption{Seasonality of the measured balanced flow as compared to that of the altimeter noise. (a)~Phase of the annual mode of the altimeter noise~$N$ measured against the phase of the mixed layer depth (MLD). (b)~Phase of the spectral slope~$-s$ compared to phase of the noise~$N$. (c)~Phase of KE2, the kinetic energy integrated over $[2k_0,4k_0]$, compared to that of the noise~$N$. (d)~Phase of KE1, the kinetic energy integrated over $[k_0,2k_0]$, compared to that of the noise~$N$.}
  \label{f:params_vs_noise}
\end{figure}

While the seasonality of the model parameters in~\eqref{eq:tidefreemodel} and of the bandpassed kinetic energy generally conform to our expectations from the winter onset of MLI, one might worry that the high-wavenumber altimeter noise has statistically significant seasonality correlated with that of the MLD, which could infect our results and create a spurious seasonal signal. We examined global maps of the Fourier power spectrum of the difference between Jason-1 and Jason-2 during the cross-calibration phase, taken as an estimate of the altimeter noise \citep{Xu2012-cm,Dibarboure2014-jd}. These are indeed red, but, as we noted above, the signal-to-noise ratio is generally of order \num{e2} or larger. Furthermore, the seasonality of he noise as measured by the phase of the annual mode has a distinct lag as compared to the seasonality of the measured spectrum of balanced ocean dynamics. In Fig.~\ref{f:params_vs_noise}, we plot a histogram of the phase of the annual mode of the noise with respect to that of the MLD and of the phases of the slope~$-s$, KE1, and KE2 relative to that of the noise. It is clear that the maximum of the annual mode of the noise occurs before that of the MLD by one to three months and several months before the maximum of the other parameters. We take this as evidence that the seasonal signals we are seeing in the slope, KE1, and KE2 are real and not simply tracking the seasonality of the altimeter noise. Note that the phase errors of the noise, slope, KE1, and KE2 are generally less than 1.5~months, so we have some confidence that these timing differences are real.

Finally, in Fig.~\ref{f:seasonal_vs_mld}e--f, we show histograms of the phases of the annual components of the amplitude $A(\phi,\lambda,t)$ of the mesoscale plateau and of the transition scale $k_0(\phi,\lambda,t)$. Here, we see much weaker seasonality, reflected in both the number of grid points with a statistically significant signal and the breadth of the peaks in these histograms. There is some tendency for the amplitude to have an annual mode that peaks six to eight months after that of the MLD and for the annual mode of the transition scale $k_0(\phi,\lambda,t)$ to have a maximum (or the wavelength to have a minimum) two to four months after that of the MLD. While a maximum in the amplitude this late compared to that of the MLD is consistent with the continuation of the inverse cascade, the breadth of the peak suggests that there are competing effects at these scales. The seasonality in $k_0$ currently lacks an explanation.

\section{Discussion}\label{S:4}

\subsection{Comparison to previous global studies}

While our focus is on seasonality, we open the discussion by comparing our results to prior studies of the spectral slope parameter~$s$. This parameter is sensitive to the details of one's analysis protocols, and it differs both between different altimetry studies and between some altimetry studies and {\it in situ} data.

\citet{Xu2012-cm} studied the along-track spectrum from Jason-2 data and analyzed the spectral slope by
\begin{enumerate}
    \item estimating the white-noise amplitude by averaging the signal over a fixed scale range (wavelengths \SIrange{25}{35}{km}) and subtracting this value from the signal at all wavelengths, and
    \item fitting a power law slope to the subtracted signal over a fixed scale range (wavelengths \SIrange{70}{250}{km}).
\end{enumerate}
There was no screening for the presence of mode-1 and mode-2 tides, and the data was averaged over the calendar year. The result of this protocol is a much gentler spectral slope, as one can see by comparing Fig.~3b of that paper and Fig.~\ref{f:slope_annual_map} here. Similar results with the same basic analysis strategy were obtained from Jason-2, Cryostat-3, and AltiKa data by \citet{Dufau2016-eg}. The resulting spectral slopes are gentle compared to both theoretical expectations \citep{Charney1971-gh,Blumen1978-pn,Callies2016-fk} and ADCP measurements \citep{Callies2013-lm,Rocha2016-kd,Shcherbina2013-pu,Callies2015-bu}.

\citet{Vergara2019-kz} studied Jason-2, AltiKa, and Sentinel-3 data, based on overlapping $\ang{15} \times \ang{15}$ patches. For the first two data sets, they estimated the noise by fitting the SSH along-track spectrum below \SI{30}{km} wavelength; for Sentinel-3, they fitted a red noise spectrum. These noise estimates were then subtracted from the data as in \citet{Xu2012-cm}. A difference from that work is that \citet{Vergara2019-kz} adjust the lower limit of the wavenumber range to which they fitted a power law to equal the local eddy length scale \citep{Eden2007-gl}. This is closer to our treatment, in which we simply fit the spectrum to the model~\eqref{eq:tidefreemodel} with a variable transition wavenumber~$k_0$. \citet{Vergara2019-kz} also studied the subtraction of internal tides \citep[following][]{Ray2016-pi,Zaron2017-sx} and found that this can increase the slopes by as much as $0.7$ in AltiKa and Sentinel-3 data.

The upshot is that \citet{Vergara2019-kz} measured spectral slopes larger than those found in \citet{Xu2012-cm}, presumably stemming in large part from the care taken in selecting the range in which they fit to a $k^{-s}$ spectrum. Comparing their Fig.~5a to our Fig.~\ref{f:slope_annual_map}a, we find a broadly similar picture, although our slopes can be steeper, as high as $k^{-6}$.

\citet{Vergara2019-kz} also studied the seasonality of the slopes by comparing spectra binned in DJF and JJA; results for the AltiKA and Jason-2 missions are shown in their Fig.~7. These windows differ by a month from the JFM and JAS windows used in our Fig~\ref{f:slope_annual_map}b--c. Again, the spatial distribution of the seasonality is broadly  similar to ours, while we find somewhat larger slopes.

It would be interesting to do a local analysis of the differences between our study and \cite{Vergara2019-kz}. For now, we compare our results directly to modeling and {\it in situ} studies.

\subsection{Comparison to regional studies} \label{ss:regional}

Our work was inspired by a number of regional studies using modeling, {\it in situ}, and altimetry results. To cross-check our results, we will compare our results to various regional studies in the existing literature. The goal here is to ensure that our results are consistent with existing work, which we find to be the case. Along the way, we encounter some interesting puzzles, particularly in the region of the South Atlantic and Southern Ocean near South Africa.

\subsubsection{Gulf Stream region} \label{sss:gs}

We start by returning to the Gulf Stream region described in Section~\ref{S:2} and comparing it to {\it in situ} data. ADCP measurements from the Oleander project \citep{rossby2019oleander} were taken from a transect that is covered by the same region, allowing a comparison to our analysis.\footnote{In doing such a comparison, we should note that balanced and unbalanced motion project differently onto SSH, so that, in general, a comparison to direct velocity measurements from ADCP data requires some care. However, in this region and in the scale range captured by Jason-2, balanced motion appears to dominate both SSH and velocities.}
\citet{Callies2015-bu} studied seasonality in the ADCP data by computing the along-track wavenumber spectrum binned over winter (JFM) and summer (JJA) months. This showed a flattening of the spectrum in winter months, at wavenumbers above the peak in kinetic energy (which is close to, and related to, the scale $k_0$). They also found that the kinetic energy at the peak and at lower wavenumbers does not show a seasonal signal. 

\begin{figure}[t]
  \noindent\includegraphics[scale=0.6]{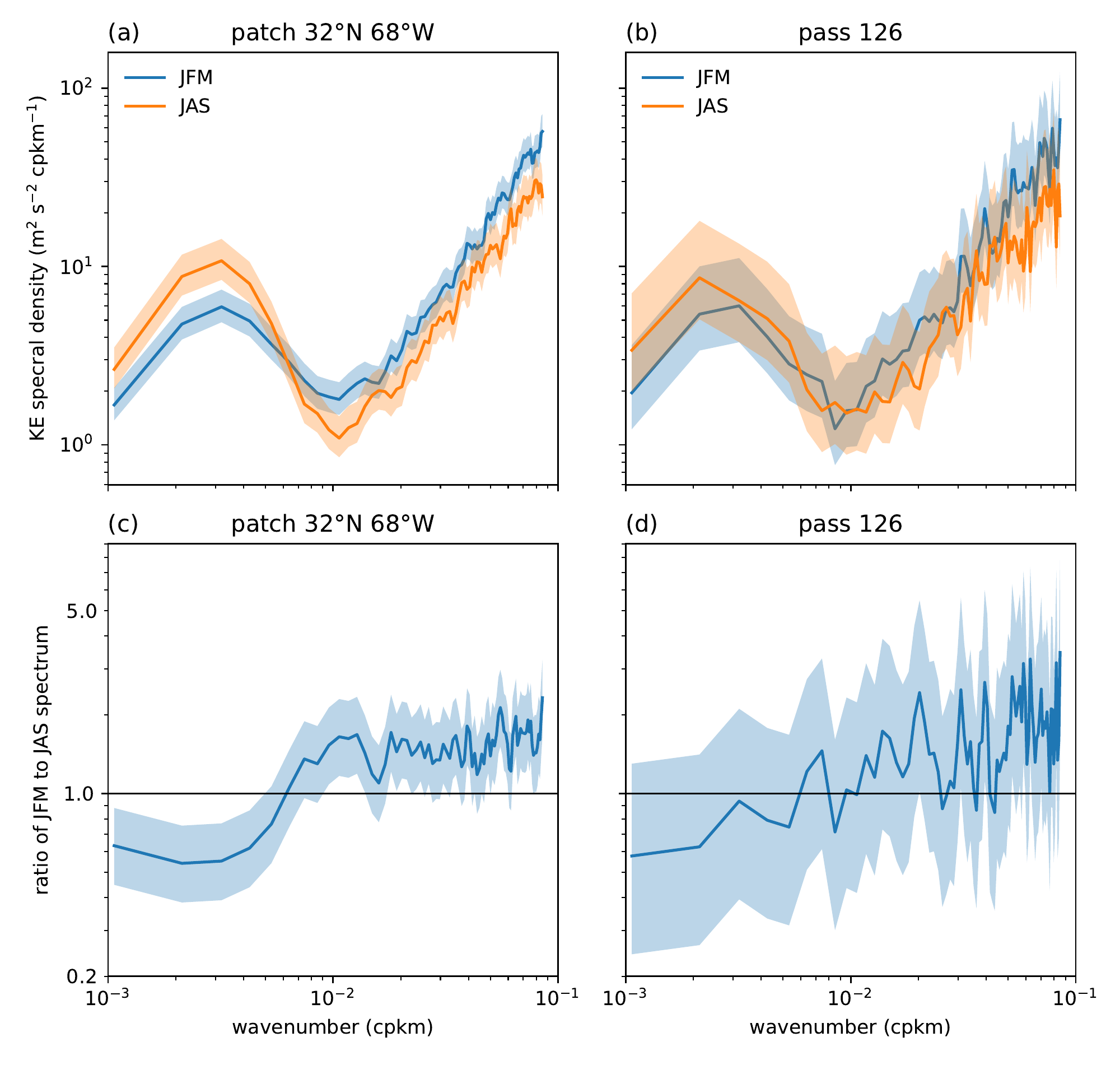}
  \caption{(a)~Power spectral density of kinetic energy in an $\ang{8} \times \ang{8}$ box in the Gulf Stream region centered at \ang{32}N \ang{68}W. Kinetic energy binned over winter months (JFM) and summer months (JJA). These spectra are taken using all passes that intersect this grid box (rather than rejecting passes with fewer than four accepted cycles in a given calendar month, as described in Appendix~A). The shading denote 95\% confidence limits. The rising signal at high wavenumbers is due to the altimeter noise, which is white in the SSH power spectrum. (c)~Ratio~$r$ of the KE power spectral density in JFM to that in JAS. The horizontal line denotes $r = 1$, and the shading denotes the 95\% confidence limit based on the assumption that $r$ has an $F$-distribution \citep[e.g.][]{thomson2014data}. (b,d)~As in (a,c) but for the kinetic energy computed from the segment of pass~126 in the same region.}
  \label{f:gulf_binned_spectra}
\end{figure}

For closer comparison to that data, we computed the KE spectra from all passes in the $\ang{8} \times \ang{8}$ region centered at \ang{32}N \ang{68}W described in Section~\ref{S:2}, binned over the same summer and winter months, with the results shown in Fig.~\ref{f:gulf_binned_spectra}a--b. As in \citet{Callies2015-bu}, we see the expected flattening of the spectrum above the peak in kinetic energy. A fit of the SSH power spectra to the model \eqref{eq:tidefreemodel} yields spectral slopes of $3.8 \pm 0.1$ in JFM and $4.8 \pm 0.2$ in JAS. These values are consistent with the power laws for KE in these months seen in the Oleander ADCP data \citep[$\sim$2 in JFM and $\sim$3 in JAS,][]{Callies2015-bu}.
Still, in the altimetry data, the power in the mesoscale peak in the kinetic energy shows seasonality, and the annual component of $A(t)$ is statistically significant.

The comparison is not yet precise, however, as our analysis covers a larger region than the single ADCP track, and the difference between our analysis and that of \citet{Callies2015-bu} could be due to spatial heterogeneity. For a closer comparison, we note that the Jason-2 flight path includes a pass (126) which runs close to the Oleander transect. The annual spectrum along this pass is known to match the ADCP spectrum once properly interpreted. \citet{Wang2010-bh} computed the along-track wavenumber spectrum of the kinetic energy from altimetry (averaged over all passes regardless of season) to the sum of the along-track wavenumber spectra of the along- and across-track velocity components. They found the signal from altimetry was notably higher. \citet{Callies2013-lm} noted that, for an isotropic spectrum dominated by balanced flow and steeper than $k^{-1}$, the along-track wavenumber spectrum of the along-track velocity is expected to be smaller than the along-track wavenumber spectrum of the across-track velocity. The KE spectrum from altimetry matches well to the power spectrum from across-track velocity taken from the ADCP \citep[see also][]{Wortham2014-ob}.\footnote{At \SI{100}{km}, the altimetry product became comparatively steep; this was due to the smoothing in the data product used in \citet{Callies2013-lm}.}

Returning to our analysis, however, we find that of the six altimeter passes that this region contains, pass 126 is not included. For the segment of pass 126 under consideration, there are large numbers of cycles with missing data points from June through August, so that this pass does not satisfy the data quality criterion we have imposed in constructing our global maps. However, binning the cycles without gaps for this pass over three months gives better statistics, and we can see the result in Fig. \ref{f:gulf_binned_spectra}c--d. The peaks in kinetic energy for JFM and JAS are the same to $95\%$ confidence level, consistent with \citet{Callies2015-bu}. Distinct from that analysis, we also find no statistically significant seasonal variation of the spectrum at any wavenumber above the noise floor, as seen in Fig. \ref{f:gulf_binned_spectra}d. The measured SSH spectral slope is $4.0 \pm 0.3$ in JFM and $3.8 \pm 0.3$ in JAS, which are statistically indistinguishable.

Further study of the three {\it accepted} passes in this region (39, 141, 217) shows that the westernmost pass (141) displays the most statistically significant seasonal variation of the low-wavenumber peak of the kinetic energy. A lesson from this comparison is that there is spatial variation of the kinetic energy spectra within the patch we are examining, which should be taken into account when comparing to more local measurements, and which deserves further investigation. One possible source of this variation is the presence of stronger zonal currents on the north side of this patch: the geostrophic velocity derived from the Aviso+ Mean Dynamic Topography product shows a strong westward recirculation current (with the eastward Gulf Stream being north of the patch we study). 

\subsubsection{Northeast Atlantic (Porcupine Abyssal Plain)}

The OSMOSIS experiment consisted of mooring and glider observations of a region centered at \ang{48.7}N \ang{16.2}W. This region is far from any strong mean currents but has a significant seasonality in the MLD \citep{Thompson2016-vq,Damerell2016-hc,Erickson2018-xk,Erickson2020-ij}. Mooring observations of kinetic energy show a seasonality at subinertial frequencies, corresponding to balanced motion, with elevated levels in winter (JAS) as opposed to summer months. More detailed frequency analysis of the mooring data, for structure functions computed at different fixed separations between \SIlist{1.3;18.7}{km}, assign a spatial scale of $\sim$\SI{10}{km} to this seasonality \citep{Callies2020-jn}. These findings are consistent with an MLI triggered in winter months that energizes the submesoscale.

\begin{figure}[t]
  \noindent\includegraphics[scale=0.55]{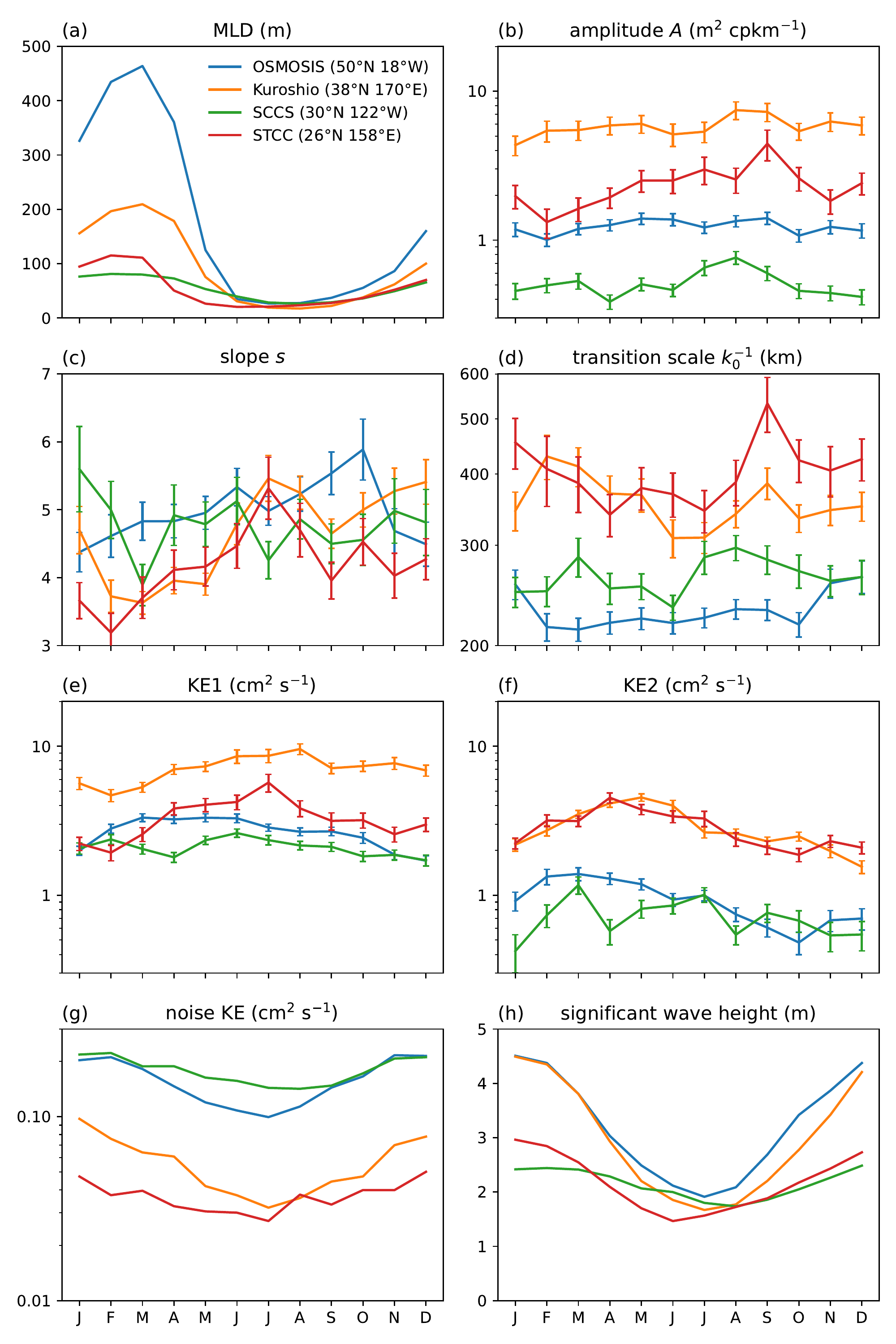}
  \caption{Monthly time series of model parameters from~\eqref{eq:tidefreemodel}, together with the MLD and SWH from climatology, in the Northern Hemisphere patches discussed in Section~\ref{S:4}.\ref{ss:regional}. (a)~Mixed layer depth from ECCO-4 climatology. (b)~Plateau amplitude~$A$. (c)~SSH spectral slope~$s$. (d)~Transition scale $k_0^{-1}$. (e)~KE1, the kinetic energy in the scale range $[k_0, 2k_0]$. (f)~KE2, the kinetic energy in the scale range $[2k_0, 4k_0]$). (g)~Altimeter noise~$N(t)$. (h)~Significant wave height from ERA5 reanalysis.}
  \label{f:nh_parameters}
\end{figure}

These observations do not yield wavenumber spectra and cover distances shorter than those resolved by altimetry. That being said, examining the parameters in the scale range covered by altimetry, we see seasonality consistent with the mooring observations (blue lines, Fig.~\ref{f:nh_parameters}). The annual mode of KE2 has a phase difference of $1.3 \pm 0.3$ months with respect to that of the  MLD, and KE1 has a phase difference of $2.8 \pm 0.3$ months with respect to that of the MLD. The minimum of the annual mode of the spectral slope $s$ differs from the maximum of the annual mode of the MLD by $-0.4 \pm 0.7$ months, so there is no significant phase shift between the two quantities.

\clearpage
%

Note that the KE2 signal (Fig.~\ref{f:nh_parameters}f) and the noise (Fig.~\ref{f:nh_parameters}g) compete in magnitude here. They have a distinct temporal structure, however, and the phase of the annual modes of the KE2 and the noise differ by three months. We have also plotted the significant wave height (SWH) from ERA5 climatology \citep{Hersbach:2020}, and it tracks the altimetry noise. (These are presumably somewhat degenerate since the ERA5 SWH is also derived from altimetry.) The annual modes of the bandpassed kinetic energies and the slope are all nonzero at a $>$99\% confidence level. There is thus good reason to believe that the seasonality we observe in the balanced motion is not simply a projection of the seasonality of the noise.

\subsubsection{Southern Ocean near Drake Passage}

\citet{Rocha2016-or} examined ADCP data, satellite altimetry, and LLC4320 model output for mesoscale and submesoscale kinetic energy in the Drake Passage region. Based on the ADCP data, they reported no statistically significant seasonality in the \SIrange{10}{500}{km} range of the kinetic energy spectrum. 

This is a complex area with strong jets and winds, which may complicate extracting information about the operation of MLI in this region. Jason-2 data is comparatively sparse. The grid box centered at \ang{58}S \ang{62}W, which includes the region studied by \citet{Rocha2016-or}, contains a single pass (35) that passes our data quality criteria; this pass runs from \ang{62.0}S \ang{72.3}W to \ang{56.8}S \ang{59.4}W and intercepts the ship tracks studied by \citet{Rocha2016-or}. (Technically this pass runs outside of the box: see Appendix~A for details of how passes are chosen and data is extracted.)

\begin{figure}[t]
  \noindent\includegraphics[scale=0.55]{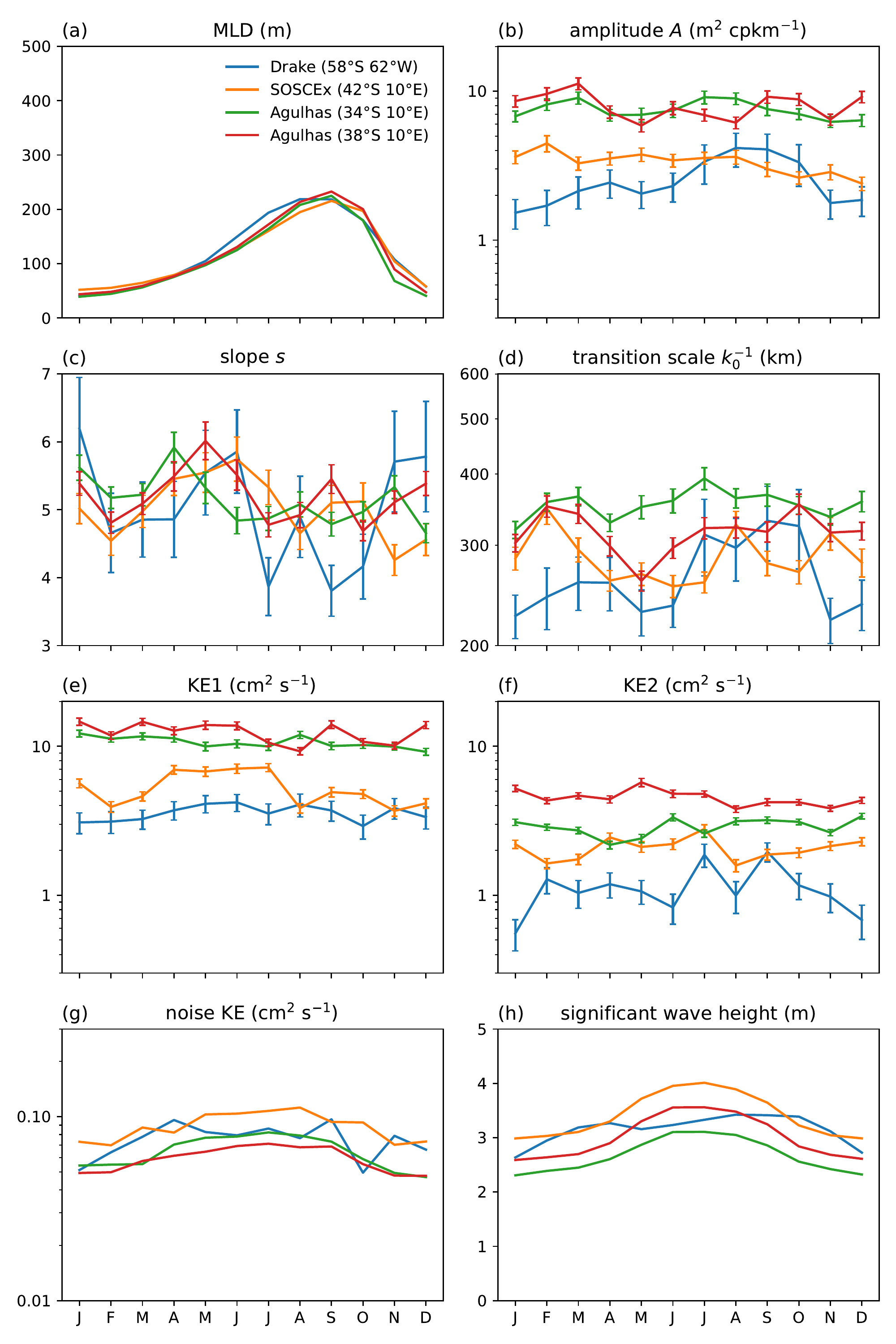}
  \caption{Monthly time series of model parameters from~\eqref{eq:tidefreemodel}, together with the MLD and SWH from climatology, in the Southern Hemisphere patches discussed in Section~\ref{S:4}.\ref{ss:regional}. (a)~Mixed layer depth from ECCO-4 climatology. (b)~Plateau amplitude~$A$. (c)~SSH spectral slope~$s$. (d)~Transition scale $k_0^{-1}$. (e)~KE1, the kinetic energy in the scale range $[k_0, 2k_0]$. (f)~KE2, the kinetic energy in the scale range $[2k_0, 4k_0]$). (g)~Altimeter noise~$N(t)$. (h)~Significant wave height from ERA5 reanalysis.}
  \label{f:sh_parameters}
\end{figure}

A time series for the parameters in model~\eqref{eq:tidefreemodel} is shown in Fig.~\ref{f:sh_parameters}. There is statistically significant seasonality above a 99\% confidence level in the annual mode of KE2 and above a 95\% confidence level in the annual mode of the slope (but less than 95\% confidence level for KE1). The phase difference between the annual modes of the slope and the MLD is $-0.1 \pm 1.3$; the phase difference between the annual modes of KE2 and the MLD is $-0.5 \pm 0.9$; both are consistent with zero. Note also that KE2 and $-s$ closely track the significant wave height; the phase differences of the annual modes of these with respect to the SWH are $0.2$ and $0.6$, respectively with the same errors as above, and so consistent with zero phase difference.
Similarly, the phase difference of the slope $-s$ and KE2 with respect to the altimeter noise are $1.8 \pm 1.3$ and $1.4 \pm 1.0$. These are all within $2\sigma$ of zero.

\clearpage

Since we have few passes in a given patch and the region is heterogeneous, we also examine a grid box centered \ang{2} farther east, which similarly contains only one pass (111) accepted by our criteria. This pass also intercepts the ship tracks studied in \citet{Rocha2016-or}, and runs close to the region identified in Fig.~1 of that paper as the Polar Front. The seasonality for KE1 and the slope are significant above a 99\% confidence limit, and for KE2 above a 95\% confidence limit. For the slope $-s$, KE1, and KE2, the phase differences with respect to the MLD are $9.5 \pm 1.0$, $4.1 \pm 0.6$, and $3.9 \pm 1.4$, respectively. Relative to the SWH, the phases of the slope $-s$, KE1, and KE2 are $-1.7$, $4.9$, and $4.7$, respectively, with the same error bars as above. With respect to the noise, they are $5.2 \pm 1.2$, $-0.2 \pm 0.9$, and $-0.4 \pm 1.5$. These last two are consistent with zero phase difference. 

If we instead study the spectra for pass~35 binned over summer (JFM) and winter (JAS) months, we find that the spectra are equivalent wavenumber by wavenumber within the 95\% confidence range. If we fit the binned data to~\eqref{eq:tidefreemodel}, however, we find spectral slopes $s = 5.1 \pm 0.4$ in the summer and $4.2 \pm 0.2$ in the winter, consistent at greater than $2\sigma$ with wintertime enhancement of the submesoscale. This apparent lack of seasonality in the binned spectra is consistent with \citet{Rocha2016-or} (although they do not specify their binning of ADCP data). Note that the slopes measured in \citet{Rocha2016-or} from ADCP and AltiKA data are $s\lesssim 5$, consistent with our results for the binned spectrum.

In summary, our data are consistent with the results of \cite{Rocha2016-or}; the apparent lack of seasonality seems to arise from binning over summer and winter months. The seasonality in the slope and kinetic energy does not quite fit the picture we find elsewhere. However, there is considerable spatial heterogeneity in this region, and the pass segments studied intersect regions with strong zonal currents.
Furthermore, the timing of the seasonality is closer to that of either the SWH (for pass 35) or the altimeter noise (for pass 111). It would be worthwhile to perform some dedicated analysis here.

\subsubsection{California Current System}

\citet{Chereskin2019-dd} studied the southern California Current System using ADCP transects, satellite altimetry, and output from the LLC4320 simulation. While model results showed a spectral slope for balanced motion, as extracted by daily averaging, that is gentler in the winter months, the ADCP data and the hourly averaged model show no statistically significant seasonality at wavelengths shorter than \SI{100}{km}, while ADCP data sees a wintertime enhancement of kinetic energy at wavelengths longer than \SI{100}{km}. \citet{Chereskin2019-dd} also studied seasonality from Jason-1 and Jason-2 data and found no statistically significant signal from data binned over spring and fall months. The authors provide evidence that the lack of seasonality in the model fields is due to the superinertial internal gravity wave signal peaking in the summer months, a signal strong enough to cancel the seasonality in the balanced motion. The difference between the model and the ADCP data may be in part because the model produces a larger signal in wave motion relative to balanced motion, as compared to the {\it in situ} data.

Our satellite data shows weak seasonality consistent with the ADCP results (Fig.~\ref{f:nh_parameters}, green lines), although submesoscale seasonality may be obscured by noise in the altimetry data, as it is by wave motions in the ADCP data. The $\ang{8} \times \ang{8}$ box centered at \ang{30}N \ang{122}W covers this region fairly well and contains three altimetry passes. The slope and the transition scale $k_0$ show no statistically significant seasonality. Inspection of the annual spectrum reveals that the noise dominates the signal at wavelengths below \SI{100}{km} (based on computing the ratio of the annual spectrum to the fitted noise and asking where the ratio becomes unity at 95\%~confidence), while $k_0^{-1} = \SI{260}{km}$. Thus, it is not surprising that the slope seasonality is at best hidden. Nonetheless, the amplitude $A$ and the bandpassed kinetic energies have nonzero annual modes at $99\%$~confidence levels. The phase relative to that of the MLD is $3.2 \pm 0.8$ months for KE2 and $3.6\pm 0.7$~months for KE1; the annual mode of the amplitude $A$ has a phase of $5.5 \pm 0.7$~months with respect to the MLD. Note that despite the KE2 signal being below the noise, as extracted from the model, it has a distinct seasonality: the annual modes of KE1 and KE2 have phase differences with respect to the noise of $4.5 \pm 0.7$ and $4.1 \pm 0.8$~months, respectively. 

Finally, we note that the measured slope in the annually averaged spectrum is $5.0 \pm 0.2$. This is consistent with the model results in \citet{Chereskin2019-dd} with the high-frequency motions filtered out. Again, we note that the part of the measured spectrum controlled by $s$ covers a factor of \num{2.6} in the wavenumber range. In and of itself, the data is not strong evidence for power law scaling, but the consistency with model results is reassuring.

\subsubsection{North Pacific Subtropical Counter Current}

\citet{Qiu2014-hn} studied the North Pacific Subtropical Counter Current (STCC), a region with strong mesoscale eddy activity roughly bounded by \SIrange{18}{28}{\degree}N and \SIrange{135}{162}{\degree}E, using $1/\ang{30}$ model output and the Aviso+ gridded altimetry product. In the model, they found evidence for an inverse cascade proceeding in time through the spring: submesoscale KE (wavelengths $<$\SI{100}{km}) peaked before mesoscale KE. They also studied the spectral energy flux using the Aviso+ gridded data; however, the large spacing between ground tracks makes interpreting this calculation challenging \citep{Arbic2013-ng}.
  
Our results in this region are consistent with an MLI triggered in the winter months, inducing an inverse cascade (Fig.~\ref{f:nh_parameters}, red lines). All displayed parameters have annual modes at a $>$99\% confidence level. The annual mode of the high-wavenumber KE2 band has a maximum $2.9 \pm 0.3$~months after the maximum of the annual mode of the MLD; that for the lower-wavenumber KE1 band has a maximum $5.1 \pm 0.4$~months after that of the MLD; the annual mode of the spectral slope has a minimum $0.0 \pm 0.6$~months after the maximum of the annual mode of the MLD.

\subsubsection{Kuroshio extension region}

Following \citet{Sasaki2014-cq} and \citet{Rocha2016-kd}, we study the Kuroshio extension region, where both mesoscale and submesoscale activity is strong. A casual glance at the maps in Fig.~\ref{f:slopemap} indicates that the seasonal dynamics here are consistent with an MLI triggered in the winter, followed by an inverse cascade. The modeling results of 
\citet{Sasaki2014-cq} and \citet{Rocha2016-kd} were also consistent with the submesoscale powered by an MLI in the winter months.

The orange lines in Fig.~\ref{f:nh_parameters} show the time series for quantities characterizing the balanced motion for a specific grid box in this region. There is no statistically significant (at 95\% confidence) annual mode of the amplitude $A$ of  the mesoscale plateau, consistent with \citet{Sasaki2014-cq} and \citet{Rocha2016-kd}. The bandpassed kinetic energies KE1 and KE2, however, as well as the spectral slope~$s$, all have annual modes consistent with the MLI mechanism. The annual mode of the slope has a minimum $1.1 \pm 0.4$~months after the maximum of he annual mode of the MLD, the annual mode of KE2 has a maximum $2.6 \pm 0.2$~months after that of the MLD, and the annual mode of KE1 has a maximum $5.4 \pm 0.4$~months after that of the MLD.


There is some variation in this region. For example, in the grid box centered at \ang{38}N \ang{162}E, \ang{8}~west of the region characterized by the orange lines in Fig.~\ref{f:nh_parameters}, the amplitude and KE1 signals have statistically significant annual modes that peak $7.8 \pm 0.4$~months and $7.6 \pm 0.4$~months, respectively, after the maximum of the annual mode of the MLD; the annual mode of the slope has a minimum $1.6 \pm .5$ months after that of the MLD; and the annual mode of KE2 has a maximum $1.7 \pm .4$ months after that of he MLD. That is to say, unlike the patch discussed above, $A$ has significant variance, and KE1 peaks more than 2~months later. Nonetheless, the seasonality still appears to be broadly consistent with submesoscale energy being activated by MLI.

The modeling results of \citet{Sasaki2014-cq} showed a spectral slope of the kinetic energy of $-2$ in March and $-3$ in September, consistent with the SSH power spectral slopes estimated here for those months (Fig.~\ref{f:nh_parameters}). 

\subsubsection{Atlantic sector of the Southern Ocean and Agulhas Current region}

Examination of Fig.~\ref{f:slopemap} shows that the Southern Ocean south and west of the African continent has seasonality distinct from other regions of the globe, at odds with the story of MLIs in winter injecting energy and initiating an inverse cascade. Without offering an explanation of this observation, we look in detail at specific sub-regions covered by other studies.

\citet{Du_Plessis2019-yb} examined glider data taken near \ang{43}S \ang{8}E to study the seasonality of submesoscale processes at \SIrange{1}{10}{km} in the mixed layer, specifically the depth and restratification of the mixed layer. They found enhanced lateral buoyancy gradients in the summer months and a delayed onset of restratification following surface warming due to strong westerly winds. This study took place at much smaller scales than our data can probe and has no information regarding KE.  The green lines in Fig.~\ref{f:sh_parameters} nevertheless shows the seasonality of the SSH power spectrum in an $\ang{8} \times \ang{8}$ region that includes the location of these observations. One can see peaks in both bands of KE that correlate with peaks in the noise and the significant wave height. The spectral slope becomes more gentle after the climatological mixed layer deepens. The annual mode of the kinetic energy in the KE2 band is not statistically significant by our criteria. For the KE1 band, the annual mode peaks $9.2 \pm 0.3$~months after the annual mode of the MLD, while the slope has a minimum $3.3 \pm 0.6$~months after that of the MLD in August. Close examination of the time series in Fig.~\ref{f:sh_parameters} shows a peak in KE2 in December to January, followed by a peak in KE1 beginning in April. The timing of these peaks and of the slope is somewhat delayed compared to the phase differences we find elsewhere. We leave to the future a study of whether these observations are related to the upscale cascade triggered by MLI, and whether other local dynamics might help explain the observed monthly variation of the SSH power spectrum.

\citet{Schubert2020-wm} directly investigated the submesoscale cascade via modeling studies of two parts of the Agulhas region. The ``ring path'' region, into which Agulhas rings propagate, is west of the southern African continent. Here, they found that the roll-off of the KE spectrum becomes more gentle in winter months at scales below \SI{80}{km}, and this seasonality disappears when MLIs are not resolved. The ``subgyre'' region, east of the continent, sits north of the Agulhas Return Current and east of the Agulhas Current proper. Here, \citet{Schubert2020-wm} also found evidence of enhanced upscale kinetic energy flux in the winter months. 

Examining Fig.~\ref{f:slopemap}, we can see that in the ``ring path'' region, the timing of the bandpassed kinetic energy is closer to that expected from the MLI mechanism, but this changes as one moves farther south. In Fig.~\ref{f:sh_parameters}, we show the monthly time series of the parameters in~\eqref{eq:tidefreemodel} for an $\ang{8} \times \ang{8}$ region centered at \ang{34}S \ang{10}E, which displays roughly the same seasonality. The annual mode of the KE2 bandpassed kinetic energy hitting its maximum $2.2 \pm 0.5$~months after that of the MLD, while the annual mode of the KE1 band peaks $7.1\pm 0.3$~months after that of the MLD. The annual mode of the slope hits its minimum $0.9\pm 0.7$~months after the maximum of the annual mode of the MLD. Moving \ang{4}~south, however, changes the story considerably, as we can see in Fig.~\ref{f:sh_parameters} (red lines). Here, the annual mode of the KE2 signal peaks at $8.0 \pm 0.6$~months after that of the MLD; the annual mode of the KE1 signal peaks $6.7 \pm 0.5$~months after that of the MLD. Only the slope behaves closer to expectations, with the annual mode reaching its minimum $2.2 \pm 1.0$~months after the maximum of the annual mode of the MLD. 

\begin{figure}[t]
  \noindent\includegraphics[scale=0.6]{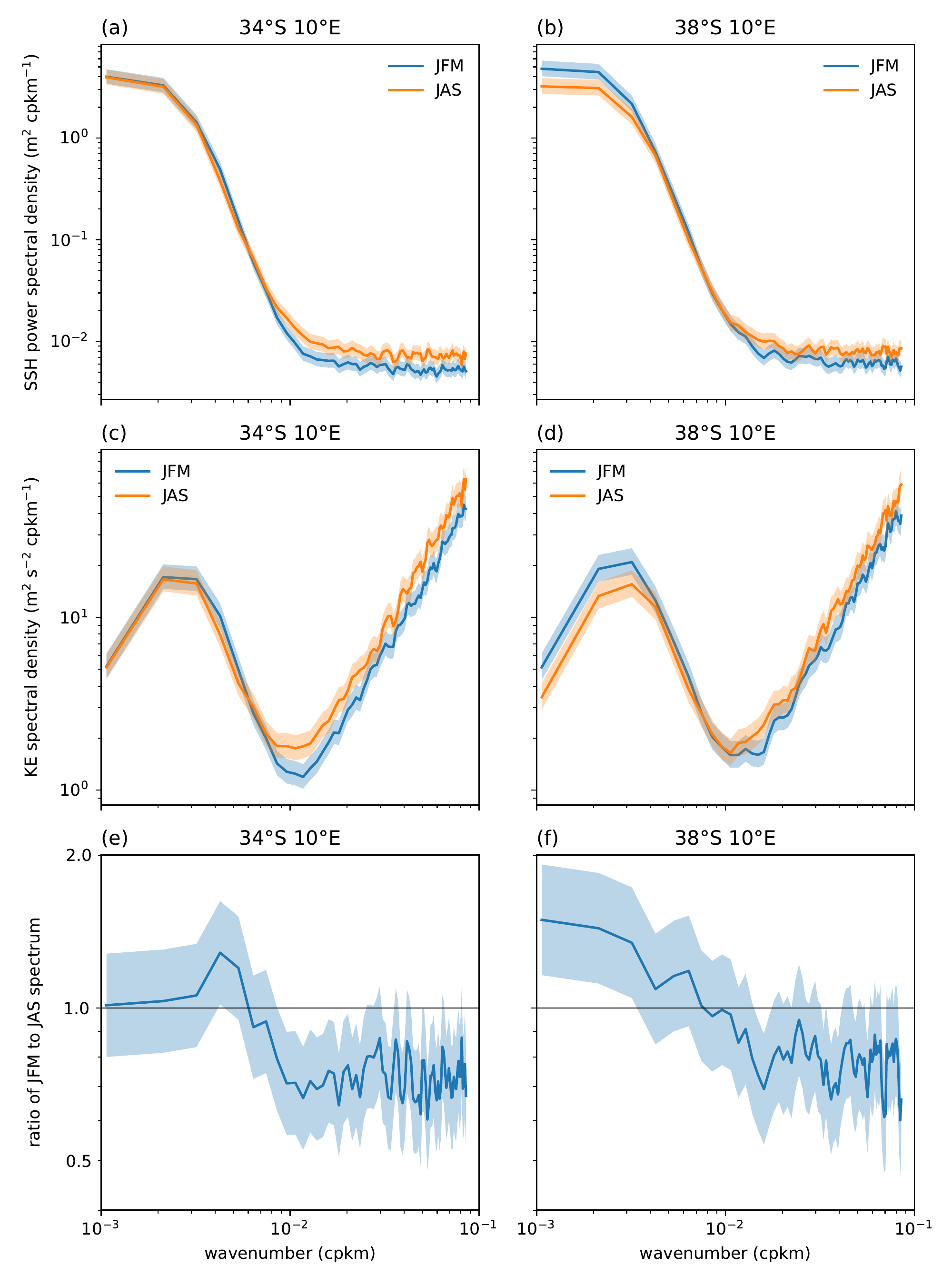}
  \caption{(a)~SSH power spectra for the $\ang{8} \times \ang{8}$ region centered at \ang{34}S \ang{10}E, inside the ``ring path'' region studied by \citet{Schubert2020-wm}. The shading denotes 95\% confidence limits for the spectra. (c)~KE for the region described in~(a); shading again denotes 95\% confidence limits. (e)~Ratio~$r$ of SSH power spectral density in JFM to that in JAS. The horizontal line denotes $r = 1$, and shading again denotes the 95\% confidence limits. (b,d,f)~As in (a,c,e) but for the $\ang{8} \times \ang{8}$ region centered at \ang{38}S \ang{10}E.}
  \label{f:ring_path_binned}
\end{figure}

To compare with \citet{Schubert2020-wm} more directly in the ``ring path'' region, we bin the SSH and KE power spectra in the summer (JFM) and winter (JAS) months, for these two boxes (Fig.~\ref{f:ring_path_binned}). For the region centered at \ang{34}S \ang{10}E, the story looks similar to Fig.~2 of \citet{Schubert2020-wm}, with little change in the wavenumbers at which the signal dominates over the noise floor. The SSH spectral slopes are $s = 5.3\pm 0.1$ in JFM and $s = 4.9 \pm 0.1$ in JAS. These slopes appear consistent with the corresponding wavenumber range in Fig.~2 of \citet{Schubert2020-wm}. For the region centered at \ang{38}S \ang{10}E, the story looks different, with the mesoscale KE peak having lower power in the winter months. 

Figure~\ref{f:slopemap} also shows some variation in and near the ``subgyre'' region, with some grid points having a timing of the bandpassed KE consistent with submesoscale energization by MLI, and others having fairly late activation of the higher-wavenumber KE2 band. This region was studied less intensively in \citet{Schubert2020-wm} and is dynamically complex, so we will leave further investigation for future work.

\section{Conclusions}\label{S:5}

Our conclusion is that wintertime energization of the submesoscale by a mixed layer instability, followed by an inverse cascade to the mesoscale over the time scale of months, is common throughout the extratropical ocean. This can be seen in the global maps in Fig.~\ref{f:slopemap} and the histograms of the phases of the slope and bandpassed kinetic energy in Fig.~\ref{f:seasonal_vs_mld}. There are some notable exceptions, particularly a statistically significant difference in seasonality in the South Atlantic and Southern Ocean south of the African continent. In other locations, such as the Drake Passage region, the effects of noise, spatial heterogeneity over the grid box, and strong zonal currents may give a competing signal. 

The distinctive features of our analysis are that we reject regions with strong tidal signals in the wavenumbers of interest, that we fit the data to a model that includes the noise and the transition scale $k_0$ (for the latter parameter, similarly to \citet{Vergara2019-kz}), and that we measure the seasonality by studying the relationship between the annual variation of the balanced dynamics as compared to that of the mixed layer depth. The spectral slopes that result match theory, models, and {\it in situ} observations, unlike many previous altimetry studies. When analyzed in specific regions, our analysis is generally compatible with {\it in situ} measurements, once we are careful to ensure that we are comparing sufficiently nearby altimeter and ship tracks and that we are using the same strategy to measure seasonality.

One lesson in favor of our analysis is that interesting dynamics happen on time scales that may be obscured by binning.
A cautionary lesson for our analysis, particularly apparent in the Gulf Stream region we study, is that combining altimetry passes in an $\ang{8}\times\ang{8}$ box can obscure spatial variation of the dynamics within that box.

We have focused on the seasonality of quantities that diagnose the dynamics above the transition wavenumber~$k_0$. A natural next question regards the seasonality in the parameters $A$ and $k_0$ in \eqref{eq:tidefreemodel}. The amplitude~$A$ displays weak seasonality, with a tendency to be out of phase with respect to the mixed layer depth. The transition scale~$k_0$, which is tied to the size of mesoscale eddies \citep[cf.\ e.g.][]{Tulloch2011-dc}, displays more distinct seasonality, with the maximum in $k_0$ (minimum in spatial scale) occurring in early spring, 2~to 4~months after the maximum of the annual mode of the mixed layer depths. The mechanisms for this seasonality is an interesting subject for further investigation.


Our hope is that this analysis provides a good template for analyzing seasonality in the data returned by the upcoming Surface Water and Ocean Topography (SWOT) mission, which should see somewhat better wavenumber resolution.

In the meantime, several further investigations are suggested by this data. One is to study the seasonality in the South Atlantic and the Southern Ocean south of the African continent, which is strikingly different from that suggested by energization via an MLI, and from the other parts of the global ocean; another is to better understand the spectra in the high-latitude Northwest Atlantic and Northeast Pacific.

\acknowledgments
We would like to thank Weiguang Wu for sharing his code. We would also like to thank and Dhruv Balwada, Jacob Steinberg for useful discussions. The work of A.L. was supported by a Provost's Research Innovation Award from Brandeis University.  This work was initiated while A.L. was at the Kavli Institute for Theoretical Physics, which is supported by National Science Foundation grant NSF PHY-1748958; it was also performed in part at the Aspen Center for Physics, which is supported by National Science Foundation grant PHY-1607611. J.C. was supported by NASA grant 80NSSC20K1140.

\datastatement
The Jason-2 satellite altimetry data was taken from the NASA Physical Oceanography Distributed Active Archive Center (PODAAC) hosted by the Jet Propulsion Laboratory, \url{https://podaac.jpl.nasa.gov/dataset/OSTM\_L2\_OST\_OGDR\_GPS}. Code used for analyzing the data can be found at \url{https://github.com/albionlawrence/sm-seasonality}. Mean Dynamic Topography data were produced and distributed by Aviso+ (\url{https://www.aviso.altimetry.fr/}), as part of the Ssalto ground processing segment. The mixed layer depth was taken from the ECCO-4 climatology \citep{ecco4-dataset,gmd-8-3071-2015,ecco_consortium_2021_4533349}. The significant wave height was taken from the ERA5 reanalysis \citep{Hersbach:2020}, downloaded from the Copernicus Climate Change Service (C3S) Climate Data Store.

\appendix[A]
\appendixtitle{Jason-2 data}\label{A:A}

We took SSHA data from the Jason-2 mission over the interval \numrange{2008}{2016} (cycles \numrange{000}{300}). The data was separated into $\ang{8}\times\ang{8}$ boxes. The centers of these boxes were spaced \ang{4} apart, between \ang{58}S and \ang{58}N, and between \ang{2}E and \ang{358}E, so that each box overlapped the other. Within each box we took segments with 160~data points that began at the southern bounding latitude (for ascending passes) or northern bounding latitude (for descending passes) and for which at least half of the segment was within the box. We rejected boxes in which the depth was less than \SI{500}{m} or for which ECCO did not provide data to compute tidal frequencies; we also rejected passes intersecting regions with depth $<$\SI{1000}{m}. The segments were sorted by calendar month and the pass in each box was rejected if there were fewer than four accepted cycles in any given month. 

The power spectrum for each pass and cycle was then taken by first subtracting the mean SSH to remove steric effects \citep{Gill1973-db} that would mix in with windowing, applying a Hann window, and taking the Fourier transform. The SSH power spectra were then averaged over all passes and cycles in a calendar month, providing a monthly series at each box; and averaged over all cycles to get a total average power spectrum.

\appendix[B]
\appendixtitle{Fitting the SSH PSD spectrum to a model function}\label{A:B}

Let $P(k,\vec{q})$ be a model of the SSH spectrum with parameters~$q^i$ assembled into~$\vec{q}$. This could be the model in \eqref{eq:tidefreemodel} or the generalization that includes semidiurnal baroclinic tides, as discussed in Appendix~C. The power spectrum $H(k)$ computed in Appendix~A is fit to the model $P(k,\vec{q})$  via nonlinear least squares, using the default algorithm in the SciPy routine \url{scipy.optimize.least_squares}. We start with the sum $\sum_k |R(k,\vec{q})|^2$ over wavenumber of the squares of the residual functions:
\be
	R_\mathrm{fit}(k,q^i) = \frac{H(k) - P(k,\vec{q})}{\sigma_k},
\ee
where the statistical weight is:
 \be
 	\sigma_k = \frac{H(k)}{\sqrt{n - 1} \sqrt{k}} = \frac{{\tilde \sigma}(k)}{\sqrt{k}},
\ee
%
with $n$ being the number of passes used to compute the spectrum (which is equal to the number of data points per wavenumber $k$.)
We then minimize this over the parameter vector~$\vec{q}$ and denote the best-fit values by~$\vec{q}_*$.
Note that the weight $\tilde\sigma$ is the standard weight if one assumes that the likelihood of the model $P(k, \vec{q})$ given the data $H(k)$ 
is well-approximated by a Gaussian near the maximum likelihood \citep[e.g.][]{thomson2014data}. We find, however, that when using this weight, the least squares algorithm seems to give too much weight to fluctuations in the transition region between the balanced flow and the altimeter noise floor, with the result that the fits to lower wavenumbers were poorly matched to the data. The additional factor of $1/\sqrt{k}$ was added to give increased weight to lower wavenumbers, which significantly improved the fit at lower wavenumbers and suppressed the effect of statistical fluctuations at the balanced motion--noise transition.

In computing the uncertainties, we use the residual   
\be
	R(k,\vec{q}) = \frac{H(k) - P(k,\vec{q})}{{\tilde \sigma}_k}\ ,
\ee
whose square properly captures the log likelihood of the model given the data. We then proceed using the standard algorithm
\citep[e.g.][]{press2007numerical} of computing the Jacobian $J_i(k,\vec{q}_*) = \partial R(k,\vec{q}_*) / \partial q^i$, corresponding to the model parameters, and the Hessian $H_{ij} = \sum_k J_i(k,\vec{q}_*) J_j(k,\vec{q}_*)$, then setting the uncertainty of parameter $q^i$ to $\sigma_{q^i} = \sqrt{(H^{-1})_{ii}}$. Uncertainties in the computed parameters $q^i$, are reported in this paper as $\pm \sigma_{q^i}$. Similarly, uncertainties in derivated quantities -- the bandpassed kinetic energies KE1 and KE2 described in Section~\ref{S:2}, and the phase and amplitudes, are computed from $\sigma_{q^i}$ using the standard formula for propagation of errors. 

\appendix[C]
\appendixtitle{Screening for the mode-1 tide}\label{A:C}

As discussed in \ref{S:2}, we reject spectra with substantial baroclinic semidiurnal tides that may dominate the SSH signal at a substantial fraction of the wavenumbers of interest (between $k_0$ and the wavenumber at which the altimeter noise dominates). Our screening procedure is as follows.

\begin{figure}[t]
  \noindent\includegraphics[scale=0.6]{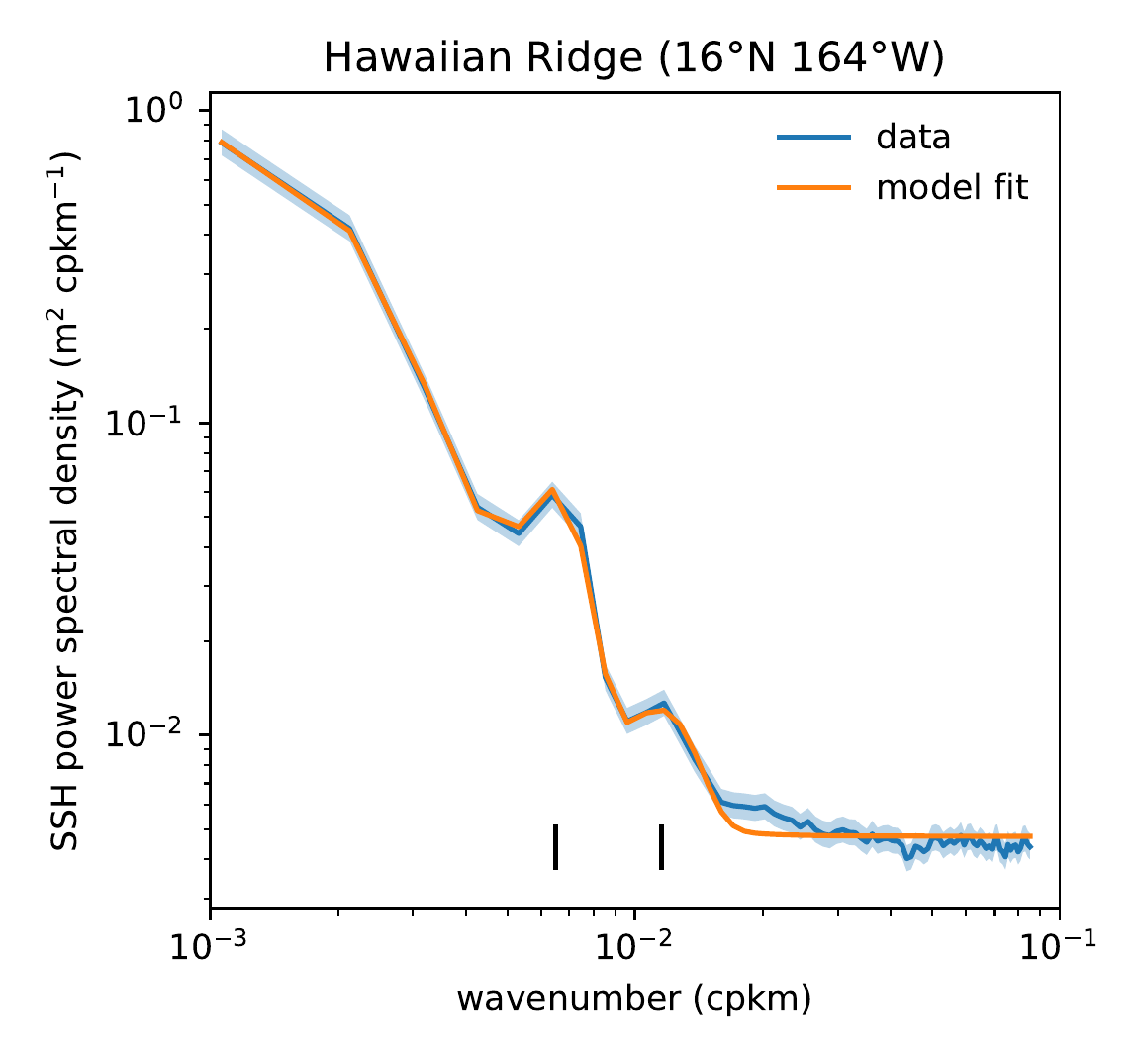}
  \caption{Along-track power spectral density of SSH in an $\ang{8} \times \ang{8}$ box centered at \ang{16}N \ang{196}E (just south of the Hawaiian Ridge), averaged over all available passes and cycles (blue line). Shading represents the 95\% confidence limits. Also shown is a fit to the model \eqref{eq:annualmodel} with the wavenumbers of the mode-1 and mode-2 semidiurnal tides computed from ECCO-4 climatology (vertical black lines). The slight excess above the white noise floor to the right of the mode-2 tidal peak is not generic across the global ocean.}
  \label{f:hawaiian_ridge}
\end{figure}

For each $\ang{8}\times \ang{8}$ patch, we first compute the total spectrum averaged over all pass segments inside the box and over all cycles.  We then fit the full spectrum to the model
\be	
	P(k) = \frac{A}{1 + (k/k_0)^s} + T_{1} e^{-\frac{(k - k_1)^2}{2 \sigma_{1}^2}}
		+ T_{2} e^{-\frac{(k - k_2)^2}{2 \sigma_{2}^2}} + N \label{eq:annualmodel}
\ee
Here, $T_{1,2}$ are the heights of the tidal peaks; $k_{1,2}$ are the wavenumbers of the mode-1 and mode-2 tides, which we compute from the ECCO-4 climatology \citep{ecco4-dataset,gmd-8-3071-2015,ecco_consortium_2021_4533349}, following the procedure in \citet{Callies2019-yf}. A plot of a spectrum with significant tides, from a region just south of the Hawaiian Ridge, also studied in \citet{Callies2019-yf}, can be seen in Fig.~\ref{f:hawaiian_ridge}. We do not completely understand the source of the broadening of these peaks, and leave that as a question for future work.

We accept grid boxes for which the observed SSH power spectrum, averaged over all months, is consistent with $T_{1,2} = 0$. We determine this by first estimating the non-tidal background as
\be
	P_\mathrm{nt}(k) = \frac{A}{1 + (k/k_{0})^{s}} + N
\ee
with $A$, $k_0$, $s$, $N$ taken from the fit to the full model. We take the power spectrum as the estimate of the variance of a signal with a Gaussian PDF at each wavenumber. Based on this, 
the data-derived spectrum~$H(k)$ is consistent with the tide-free component at a $95\%$ confidence level if it satisfies
\be
	H(k) < \frac{\chi_{0.95,m-1}}{m-1} P_\mathrm{nt}(k)
\ee
where $m$ is the sum over all pass segments and cycles within a given box, and $\chi_{0.95,m-1}$ is the upper limit of the range of a chi-square distributed variable for $m-1$ degrees of freedom that carries $95\%$ of the distribution. We take the tidal peak at $k_i$ as significant if $H(k_i)$ is inconsistent with the tide-free component at this level of confidence.

In practice we reject regions with significant mode-1 semidiurnal tidal signal, without further estimating the mode-2 signal. In surveys of the SSH power spectra in a variety of grid boxes over the global ocean, we find that if the mode-2 signal is large, the mode-1 signal is large. Furthermore, $k_2$ is typically close to the wavenumber at which the altimeter noise begins to dominate the signal, and so small mode-2 signals are difficult to distinguish from noise. Indeed, we believe that when fitting to both mode-1 and mode-2 tides, statistical fluctuations near the altimeter noise floor were often being fit as mode-2 tides, causing us to reject too many spatial regions.

\bibliographystyle{ametsocV6}
\bibliography{seasonalrefs}

\end{document}